# Halogen-Terminated Carbon Atomic Wires by Laser Ablation in Halogenated Organic Solvents: Synthesis and Characterization

Pietro Marabotti*[a,b], Simone Melesi[a], Piotr Pińkowski[c], Bartłomiej Pigulski[c], Sonia Peggiani[a], Alice Cartoceti[a], Patrick Serafini[a], Barbara Rossi[d], Valeria Russo[a], Sławomir Szafert*[c]. and Carlo Spartaco Casari*[a]

[a] Department of Energy, Micro and Nanostructured Materials Laboratory - NanoLab, Energy, Politecnico di Milano, Via Ponzio 34/3, Milano 20133, Italy

[b] Institut für Physik, Humboldt-Universität zu Berlin, 12489, Berlin, Germany

[c] Faculty of Chemistry, University of Wrocław, 14 F. Joliot-Curie, Wrocław 50-383, Poland

[d] Elettra Sincrotrone Trieste, S.S. 114 km 163.5, Basovizza, 34149, Trieste, Italy

*Corresponding authors: pietro.marabotti@polimi.it, slawomir.szafert@uwr.edu.pl, carlo.casari@polimi.it

**Abstract**

We report the synthesis of halogenated carbon atomic wires (halopolyynes) via pulsed laser ablation in liquid and their comprehensive Raman characterization. Using dichloromethane and dibromomethane-containing solutions, we produced polydisperse mixtures of monohalogenated ($HC_{2n}X$) and dihalogenated ($XC_{2n}X$) polyynes (X=Cl, Br; n=3-10). High-performance liquid chromatography enabled separation and analysis of these compounds, while chemical derivatization and mass spectrometry confirmed their molecular structures. A possible formation mechanism is proposed, involving carbon chain polymerization and termination by hydrogen and halogen atoms from atomized solvent molecules during the plasma phase. UV-Vis absorption and synchrotron-based UV Resonance Raman spectroscopy reveal that halogen terminations act as auxochromes through p-π conjugation between their lone-pair electrons and the sp-carbon backbone, extending effective conjugation length. This interaction reduces bond length alternation and moderately redshifts vibronic absorption and vibrational modes. Resonance Raman spectra show selective overtone enhancement and vibrational anharmonicity consistent with carbyne-like materials. These findings expand the synthetic approaches for halogenated carbon atomic wires and establish halopolyynes as a platform for tailoring electronic and optical properties of sp-carbon wires.

## 1. Introduction

Carbon atomic wires are appealing one-dimensional carbon materials made of sp-hybridized carbon atoms [1,2]. They represent one of the realizations of carbyne, the ideal model of a one-atom-thick linear chain of sp-carbon atoms, classified as the last allotropic form of carbon whose synthesis had remained uncharted [2]. Carbon atomic wires consist of carbyne-like chains whose properties are ruled by size confinement, i.e., chain length and terminations [1]. This flexibility originates from the high electron-phonon coupling and π-electron conjugation of the sp-carbon backbone. Scientists have developed several strategies to tailor carbon atomic wires' length and endgroups, obtaining a collection of chains characterized by tunable optical, electronic, and vibrational properties [3–7]. Thus, these compounds find many different scientific and industrial applications. For example, carbon atomic wires have been used in many prototypal applications, such as biological markers, hydrogen storage media, photovoltaic devices, and have been integrated into field-effect transistors, anti-counterfeiting labels, and supercapacitors [8–17].

Carbon atomic wires can be synthesized via chemical or physical methods. While chemical processes offer high control of the product structure and high yield, they typically rely on tailored molecular precursors and multistep reactions [6,10,18–29]. Physical methods, instead, are attractive due to their simplicity, flexibility, and scalability, and provide access to far-from-equilibrium growth regimes that are difficult to reproduce by conventional chemistry [7,30–32]. State-of-the-art techniques include supersonic cluster beam deposition [32], arc discharge [2,31,33–35], and pulsed laser ablation in liquid (PLAL) [7,36–39], in which carbon atomic wires are synthesized within a medium, usually a liquid, via a process involving a plasma phase. In PLAL, a laser is focused onto a solid target, promoting the formation of a dense plasma plume where highly out-of-equilibrium conditions facilitate the formation of sp bonds [7,36,37,40]. In particular, PLAL has been widely used to synthesize carbon atomic wires in polyyne form, i.e., with alternating triple and single bonds. PLAL experiments produce polydispersed mixtures of carbon atomic wires, with chain lengths ranging from 6 to 30 carbon atoms, various terminations ranging from H and CN to $CH_3$, and distinctive crystal structures, such as pseudocarbynes [7,41–45]. Unlike chemical methods, PLAL bypasses many of the complex steps needed to polymerize isolated chains from molecular precursors (i.e., 1 or 2 sp-carbon bonds), enabling a one-step, catalyst-free synthesis. Although PLAL-made carbon atomic wires are inherently polydispersed and cannot reach the maximum chain lengths achieved via chemical routes (30 vs 52 carbon atoms), this approach enables the exploration of formation mechanisms, termination pathways, and molecular structures unreachable through well-established and robust organic chemistry reactions [2,7,41,46].

One of the valuable advantages of PLAL is the possibility of fabricating carbon atomic wires in a wide range of solvents, potentially any. In the presence of a carbonaceous target, carbon atomic wires can also form in water, a solvent that does not supply any carbon atoms for chain polymerization [44,47–51]. Ablation of non-carbonaceous targets in organic solvents also results in carbon atomic wires formation [15,52]. Solvent parameters, such as carbon-to-hydrogen ratio, viscosity, and polarity, play a critical role in the synthesis process, enabling, for example, tuning of wire length [7,41,44,47,50,53–55]. The solvent also provides the terminating groups that stabilize the sp character of the growing chain [2,7].

This feature makes PLAL particularly suited for exploring termination chemistries that are challenging to access or stabilize via conventional synthetic routes. Based on current knowledge, carbon atomic wires produced by PLAL show hydrogen ($HC_{2n}H$), methyl ($HC_{2n}CH_3$), cyano ($HC_{2n}CN$), or dicyano ($NC_{2n}N$) terminations [42–44,54,56–58].

Among the possible terminations that remain unexplored using PLAL, halogen atoms (Cl, Br, and I) represent intriguing candidates. Indeed, chemically synthesized halopolyynes – carbon atomic wires in polyyne form capped by one or two halogen atoms – show selective reactivity and act as valuable precursors for reactions intended to tailor chain structure and properties [6]. Starting from halopolyynes, researchers have fabricated organometallic-capped wires [23–25], push-pull chains [24], products of C-C coupling reactions [59,60], polymers [61], and used them in crystal-to-crystal reactions [62]. However, known halopolyynes are usually substituted, and a general approach able to directly synthesize bare chains with only halogen terminations is still missing, mostly due to the low stability of such species. To date, longer bare halogenated carbon chains are known only for the most stable $IC_{2n}I$ series. Polyynes $IC_6I$, $IC_8I$ [63], and $IC_{10}I$ [64] were synthesized via classical wet organic synthesis, with $IC_{10}I$ being the upper limit of this approach due to explosive decomposition around 55–56 °C.

Several characterization methods have been used to detect and investigate carbon atomic wires' properties. Reversed-phase high-performance liquid chromatography exhibits unique capabilities in efficiently separating chains with different lengths and terminations, enabling size- and termination-selected analysis based on their distinct retention times, i.e., the time each compound takes to be eluted [18,41,43,44,53,65–68]. UV-Vis absorption spectra, often extracted during chromatography, provide primary data on the wires' optoelectronic properties and allow estimation of their electron-phonon coupling [3–5,38,41,41,69–73]. Raman spectroscopy remains the most powerful technique for investigating carbon atomic wires. A single Raman spectrum can provide significant insights into their vibrational, electronic, and structural properties. The fingerprint collective Raman mode of carbon atomic wires enables such comprehensive analysis, as its frequency is directly linked to bond length alternation, which is modulated by chain length and termination, charge transfer, and π-electron conjugation [4,29,46,67,73–75]. Furthermore, Raman analysis enables the quantification of vibrational anharmonicity and electron-phonon coupling [69,70]. Despite this potential, carbon atomic wires produced by PLAL are typically obtained in low concentrations, below the detection limit of conventional Raman spectroscopy. Therefore, signal enhancement is required, either by matching an electronic transition with the excitation wavelength (i.e., Resonance Raman spectroscopy) or by using plasmonic enhancement effect from noble metal nanoparticles (i.e., surface-enhanced Raman spectroscopy) [44,47,57,67,69,70,74,76–83].

Here, we present the first successful synthesis of halopolyynes by pulsed laser ablation of a graphite target immersed in halogenated organic solvents. PLAL experiments using dichloromethane or a cyclohexane–dibromomethane mixture as solvents yielded new compounds with unique retention times and UV-Vis absorption patterns, corresponding to previously unobserved monohalogenated and dihalogenated polyynes. Additional structural confirmation was provided through selective derivatization reactions with a Pd-based organometallic group and mass spectrometry measurements. We also provide a comprehensive description of the halopolyynes formation mechanism during PLAL. Their optoelectronic and vibrational properties were systematically investigated using UV-Vis absorption and synchrotron-based UV Resonance Raman (UVRR) spectroscopy, revealing the effects of halogen termination. From the Raman overtones of the wires' collective vibrational mode, we evaluate the anharmonicity of halopolyynes using our recent model [70,84]. These findings establish PLAL as a complementary route for halopolyyne synthesis, advancing the understanding of their formation mechanisms and enabling systematic studies of termination effects, while opening new possibilities to tailor carbon atomic wire structures and properties under non-equilibrium conditions.

## 2. Materials and Methods

### 2.1 Synthesis of Halopolyynes via Pulsed Laser Ablation in Liquid

The experimental setup was identical to that described in our previous work [44]. Pulsed laser ablation in liquid (PLAL) was performed using the fundamental wavelength (1064 nm) of a Nd:YAG laser (Quantel Q-smart 850), operating at 10 Hz with a pulse duration of 5 ns. The selection of 1064 nm ablation wavelength was based on previous findings showing enhanced polyyne production at longer wavelengths due to reduced optical shielding and photo-fragmentation, while favoring thermal ablation conditions that promote longer chain formation [7,85–88]. The laser beam was focused onto a graphite target (purity 99.99 %, Testbourne Ltd.) through a lens with a 200 mm focal length. The target was immersed in 5 mL of dichloromethane (DCM, HPLC grade stabilized with amylene ≥99.9% purity, Carlo Erba reagents), a 2/3 *v*/*v* solution of cyclohexane (Cyhex, ≥99.8% purity, Carlo Erba reagents) and dibromomethane (DBM, 99% purity, Sigma Aldrich), or pure isopropyl alcohol (i-PrOH, for analysis-ACS-Reag., Carbon Erba reagents). The addition of cyclohexane was necessary because DBM alone has a higher density (2.477 g/cm$^3$) than graphite (2.23 g/cm$^3$), causing the target to float. This corresponds to a liquid layer of approximately 2.5 cm above the target surface, as estimated from the solvent volume and the geometry of the vial and target. Laser pulse energy was set to 80 mJ/pulse, and the target-to-lens distance was adjusted to obtain a spot radius on the target of approximately 1.2 mm, resulting in a laser fluence of about 1.7 J/cm$^2$. The effective laser spot size was estimated using a transfer-matrix optical model that accounts for the full optical path of the laser beam, including refraction at the liquid interface and the liquid layer thickness. The model is described in detail elsewhere [77]. Each ablation was carried out for 30 minutes.

## 2.2 Purification and Concentration of Polyyne Mixtures

After ablation, the mixtures containing polyynes were transferred into acetonitrile (MeCN). This step is necessary since our reversed-phase high-performance liquid chromatography (RP-HPLC) apparatus operates in an aqueous mobile phase, while both DCM and DBM are immiscible with water, resulting in poor chromatographic analyses. Cyclohexane was used extensively as a stable nonpolar extraction solvent for polyynes, enabling their separation from undesired byproducts. Furthermore, the solutions were concentrated to enhance the polyyne concentrations, facilitating more effective characterization. The concentrated solutions were prepared from 10 ablations, obtained using the previously described parameters. The resulting 50 mL of polydispersed polyyne mixtures were purified, transferred into MeCN, following solution-specific protocols, and reduced to approximately 1 mL using a rotary evaporator (IKA Rotary Evaporators model RV8). All rotavapor processes were conducted by immersing the evaporation flask in a water bath at 40 °C, a temperature at which polyynes remain stable, and avoiding complete drying of the solutions to prevent polyynes degradation from air exposure. The same purification steps, outlined below, were applied to both as-prepared and concentrated solutions.

For ablations in DCM, the mixture was filtered using a short pad (1-2 cm) of silica gel (0.063-0.200 mm, Labkem). An equal volume of Cyhex was added, and DCM was evaporated using a rotary evaporator, transferring polyynes into the Cyhex phase. An equal volume of MeCN was then added, and Cyhex was evaporated using a rotary evaporator until 2-5 mL of polyynes in MeCN remained.

For ablations in DBM, an equal volume of dimethylformamide (DMF, HPLC grade ≥99.7% purity, Alfa Aesar) was added. The higher boiling point of DMF (153 °C) allowed the removal of DBM (97 °C) using a rotary evaporator. To remove most of the byproducts, polyynes in DMF were then extracted at room temperature with an equal volume of Cyhex and approximately three times the volume of a supersaturated solution of water and brine in a separatory funnel. Residual water in the Cyhex solution was removed with anhydrous magnesium sulfate (≥99.5% purity, Sigma-Aldrich). Finally, the Cyhex phase was mixed with an equal volume of MeCN, and the Cyhex was evaporated using a rotary evaporator until 2-5 mL of polyynes in MeCN remained.

## 2.3 High-Performance Liquid Chromatography analyses

The polyynes solutions were analyzed using an RP-HPLC (Shimadzu Prominence UFLC) equipped with a photodiode array (DAD) UV-Vis spectrophotometer, a fraction collector (FRC-10A), and a C18 column (Phenomenex Luna 3 µm C18(2) 100 Å, LC Column 150 x 4.6 mm). The gradient method employed is described in another work [77].

## 2.4 Derivatization and Mass Spectroscopy

High-resolution mass spectra were acquired using a Bruker qTOF compact equipped with an ESI ion source. Samples were derivatized as follows: 1.5 mL of MeCN fraction obtained in HPLC was mixed with 1.5 mL of DCM and an excess of water. The vial was tapped and shaken vigorously. The resulting organic phase was pipetted out to the fresh vial, and a small amount of anhydrous $MgSO_4$ was added. After filtration, we added 50 μL of a freshly prepared solution of $Pd(PPh_3)_4$, obtained by dissolving 5 mg of $Pd(PPh_3)_4$ in 30 mL of DCM. Right before the analysis, the solution was diluted using MeOH to increase ionization efficiency.

### 2.5 UV Resonance Raman Spectroscopy

We collected UV Resonance Raman (UVRR) spectra using the synchrotron-based UVRR set-up available at the BL10.2-IUVS beamline of Elettra Sincrotrone Trieste (Italy) [89]. Resonance Raman spectroscopy was performed employing the experimental method and parameters described in other works [69,70]. We employed different excitation wavelengths in the deep UV range (see Table S.3 in the SI). The wavelength tunability is granted by finely varying the aperture of the undulator gap, providing synchrotron radiation (SR) ranging from about 200 to 272 nm. The outcoming SR light was further monochromatized using a 750 mm focal length spectrometer (Acton Princeton) equipped with holographic gratings with 1800, 2400, and 3600 grooves/mm. We excited polyynes at the correspondence of their vibronic lines associated with the $|m\rangle_g \rightarrow |k\rangle_e$ transition. Given the wavelength-limitation of the tunable UV source, we excited $HC_{2n}X$ (n=4,5; X=Cl, Br) and $XC_{2n}X$ (n=4, 5; X=Cl, Br) at their strongest fundamental $|0\rangle_g \rightarrow |0\rangle_e$ (see Tables S.1 and S.3 in the SI), namely between the low-lying ($|m = 0\rangle$) vibrational level of the ground ($g$) electronic state and the lowest-energy ($|k = 0\rangle$) vibrational level of the first dipole-allowed excited ($e$) electronic state. $HC_{12}X$ (X=Cl, Br) and $HC_{14}X$ (X=Cl, Br) were measured at their $|0\rangle_g \rightarrow |1\rangle_e$ and $|0\rangle_g \rightarrow |2\rangle_e$ transitions, respectively (see Table S.3 in the SI).

### 2.6 Density Functional Theory Calculations

Time-dependent density functional theory (TD-DFT) calculations of vibronic UV-Vis spectra were carried out at the CAM-B3LYP/cc-pVTZ level of theory using Gaussian09 [90] and home-made programs [33,91], as described in our previous works [44]. The calculated spectra were rescaled by a factor of 1.1067 to make the TD-DFT calculated and experimental spectra of hydrogen-capped polyynes match.

Optimization of molecular structures and simulation of Raman spectra were conducted at the PBE0/cc-pVTZ level of theory using Gaussian 16, Revision C.01 [92]. Raman activities were converted to intensities using Multiwfn 3.8 [93] according to the known procedure [94]. The same scale factor, 0.9407, was applied to all Raman spectra. The scale factor is an average deviation between frequencies of experimental and DFT-calculated Raman bands of all analyzed halopolyynes.

## 3. Results and Discussion

### 3.1 Synthesis of Halopolyynes through PLAL in Halogenated Organic Solvents

We performed ablation experiments on a graphite target immersed in halogenated organic solvents, i.e., dichloromethane (DCM) and a mixture of cyclohexane (Cyhex) and dibromomethane (DBM) (see the Materials and Methods section), to produce halogen-terminated polyynes. According to our knowledge, this is the first time this technique has been used to synthesize such a class of polyynes. To gain a thorough understanding of the ablation products, we performed high-performance liquid chromatography (HPLC) analyses on the ablated solutions (DCM and Cyhex/DBM mixtures), transferred into acetonitrile (MeCN), using the procedure described in the Materials and Methods section. We compared these data to an HPLC analysis of a mixture of polyynes in isopropyl alcohol (i-PrOH), obtained under the same ablation conditions.

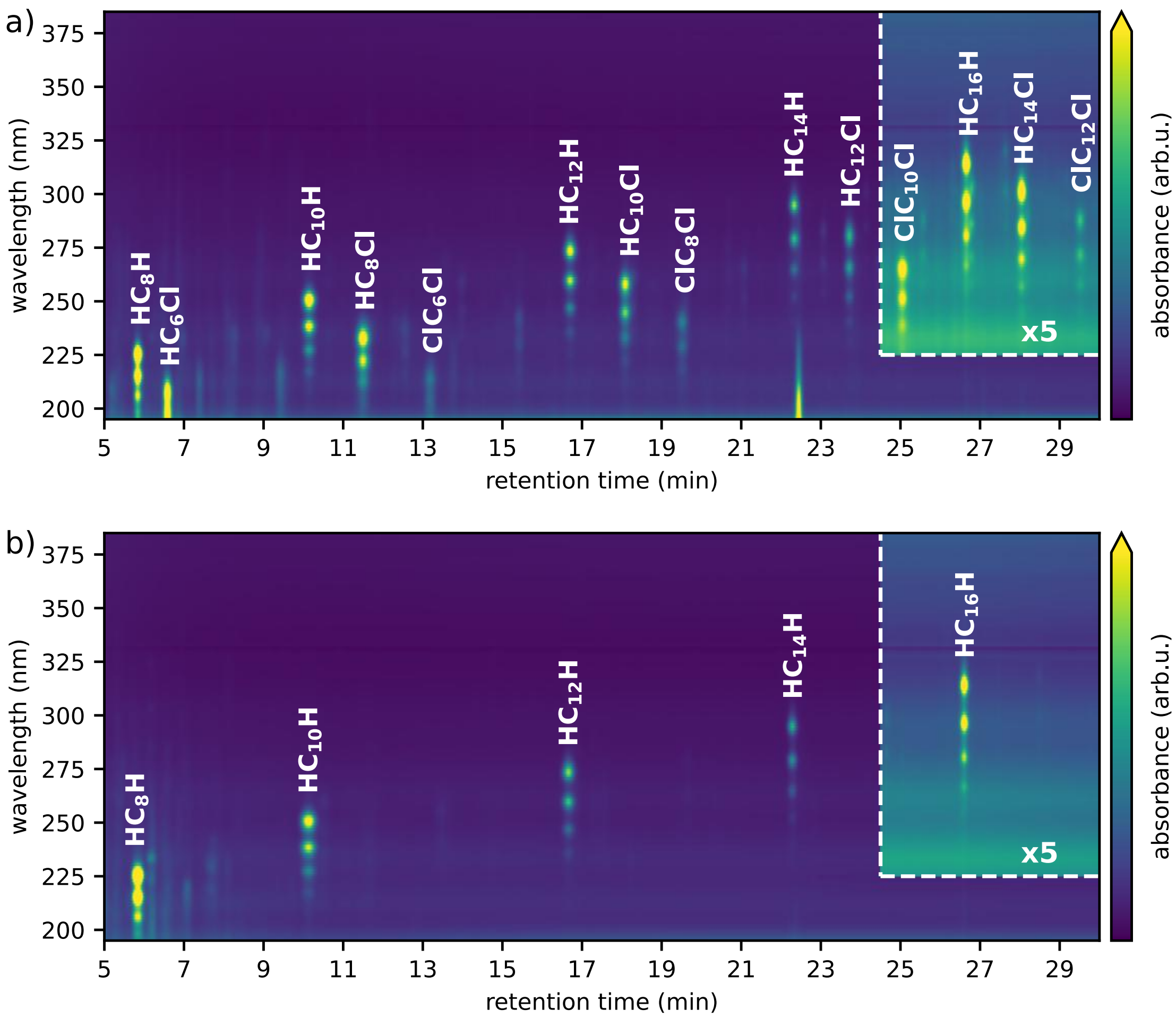

**Figure 1** Portions of the HPLC analyses of a) an ablation in DCM (after transfer in MeCN; see the Materials and Methods section) and b) an ablation in i-PrOH (directly injected). The injection volumes were 15 µL. These 2D representations show the absorbance (color scale) as a function of retention time and wavelength, extracted from the DAD coupled to the HPLC system.

Fig. 1a shows part of the HPLC analysis of a mixture of polyynes produced by ablation in DCM and transferred into MeCN for analysis (see the Materials and Methods section). Similar HPLC experiments were conducted for ablations in pure i-PrOH and Cyhex/DBM solution (the latter transferred into MeCN before HPLC analysis, see the Materials and Methods section), as reported in Fig. 1b and Fig. S.1 in the SI, respectively. Comparing these analyses, we observe new and intense chromatographic peaks in the mixtures synthesized in halogenated organic solvents, which are absent from ablations in i-PrOH. We grouped these new compounds into two distinct classes based on their retention times and similar redshifts in their vibronic progressions. Given that halogenated solvents can dissociate to provide halogen terminations, we assigned these compounds as (mono)chloro- ($HC_{2n}Cl$) and dichloro-capped ($ClC_{2n}Cl$) polyynes from the DCM ablation (Fig. 1a), and (mono)bromo- ($HC_{2n}Br$) and dibromo-capped ($BrC_{2n}Br$) polyynes from the Cyhex/DBM ablation (Fig. S.1, SI).

Our assignments are based on empirical rules developed from prior studies on hydrogen-, methyl-, and cyano-polyynes [43,44,70,95] which relate substitution of an H termination with a small functional group to a redshift in the vibronic spectrum. The extent of this redshift depends on the substituent and the interaction mechanism: for instance, a methyl endgroup induces a modest redshift (~2-4 nm or ~26-90 meV) through hyperconjugation, while a cyano termination yields a more substantial shift (~10-19 nm or ~110-466 meV) via π-π conjugation. For halogen terminations, we expect a non-negligible redshift as halogens act as auxochromes. Through p-π conjugation between their lone-pair electrons and the sp-carbon backbone π-electron system, they extend the effective conjugation length and consequently redshift absorption features. This auxochromic effect is well-established in other conjugated systems, such as halogen-substituted benzenes [96]. This redshift is typically smaller than that induced by the addition of full triple bonds (e.g., 25 nm or 548 meV passing from $HC_8H$ to $HC_{10}H$) since termination effects are generally localized. However, some terminations, such as push-pull [10,97,98] or bulky endgroups [3,4,18,99,100], can significantly perturb the sp-carbon backbone. Additionally, halogen substitution reduces the hydrophilicity of the chains, increasing their affinity for the nonpolar stationary phase of our HPLC system, thus resulting in longer retention times compared to hydrogen-, methyl-, or cyano-capped polyynes of similar length.

To illustrate our logic, we focus on the assignment of chloro-capped chains with 8 sp-carbon atoms, ($HC_8Cl$ and $ClC_8Cl$) in Fig. 1a, comparing them to their hydrogenated counterpart, $HC_8H$. $HC_8H$ has its strongest vibronic absorption peak, $\lambda_{max}$ (or optical gap, $E_{opt}$, where $E_{opt}$ is the equivalent energy of $\lambda_{max}$ in eV), at 226 nm ($E_{opt}$~5.49 eV) and is eluted at ca. 6 min (see Fig. 1). An unknown compound eluted at 6.5 min, absent in the i-PrOH mixture (Fig. 1b), has a $\lambda_{max}$ at 207 nm ($E_{opt}$~5.99 eV), blueshifted by 19 nm (~0.50 eV) compared to $HC_8H$. This suggests a shorter halogenated chain, possibly $HC_6Cl$ ($HC_6H$ has a $\lambda_{max}$ at 199 nm and $E_{opt}$ at 6.23 eV). Another unknown compound, eluted after $HC_{10}H$ ($\lambda_{max}$ = 251 nm and $E_{opt}$~4.94 eV), has a $\lambda_{max}$ at 233 nm ($E_{opt}$~5.32 eV). While this is insufficient for a 10 sp-carbon chain, this aligns with the expected redshift for $HC_8Cl$, supported by its delayed elution (11.5 min). We therefore assign this compound to $HC_8Cl$, and the earlier eluted species (6.5 min) as $HC_6Cl$.

Following this reasoning, we can predict that the other $HC_{2n}Cl$ will exhibit a redshifted $\lambda_{max}$ of ~4.5-7.5 nm (redshift in $E_{opt}$ ranging from 45 to 224 meV) relative to their hydrogenated analogs and will elute after hydrogenated polyynes with an additional triple bond (i.e., $HC_{2(n+1)}H$). This pattern also holds for $HC_{2n}Br$ species from Cyhex/DBM ablations (Fig. S.1 in the SI). Table S.1 in the SI reports the exact retention times and vibronic absorption peaks for all $HC_{2n}Cl$ and $HC_{2n}Br$.

Accordingly, the compound eluted after $HC_8Cl$ in Fig. 1a with a $\lambda_{max}$ of 214 nm ($E_{opt}$~5.79 eV) likely corresponds to a 6 sp-carbon atom wire. The $ClC_8Cl$ species was identified by elution after $HC_{12}H$ (16.7 min, $\lambda_{max}$ = 273 nm and $E_{opt}$~4.54 eV) and $HC_{10}Cl$ (18.1 min, $\lambda_{max}$ = 258 nm and $E_{opt}$~4.81 eV), specifically at 19.5 min with a $\lambda_{max}$ of 241 nm ($E_{opt}$~5.14 eV), redshifted by 15 nm (~0.34 eV) from $HC_8H$. The almost doubled redshift and the delayed elution support its assignment as $ClC_8Cl$. Similar logic supports assignments of $ClC_{2n}Cl$ and $BrC_{2n}Br$ molecules, as shown in Fig. 1a and Fig. S.1 and Table S.1 in the SI.

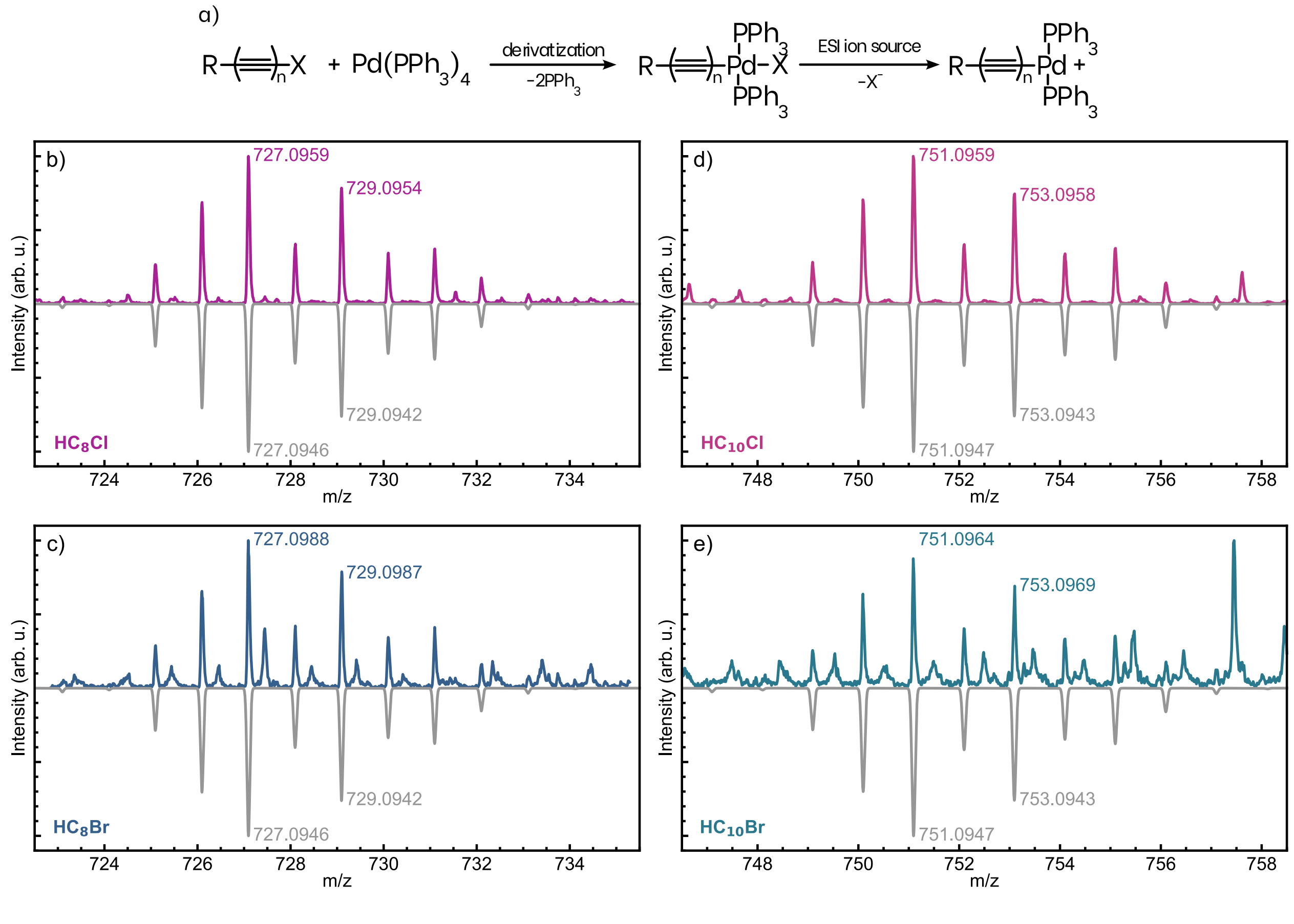

**Figure 2** a) Derivatization of halopolyynes using $Pd(PPh_3)_4$. Experimental (colored) and simulated (gray) mass spectra of b) $[HC_8(Pd(PPh_3))Cl - Cl^-]^+$, c) $[HC_8(Pd(PPh_3))Br - Br^-]^+$, d) $[HC_{10}(Pd(PPh_3))Cl - Cl^-]^+$, and e) $[HC_{10}(Pd(PPh_3))Br - Br^-]^+$.

High-resolution mass spectrometry (HRMS) measurements were carried out to confirm the chemical structure of halopolyynes. However, all direct attempts to observe such species using APCI (atmospheric-pressure chemical ionization) and ESI (electrospray ionization) ion sources failed, probably due to insufficient ionization efficiency or instability of halopolyynes in the ion source. To address this, we leveraged their known reactivity for derivatization: halopolyynes readily undergo oxidative addition with $Pd(PPh_3)_4$, forming $R(C{\equiv}C)_nPd(PPh_3)_2X$ species, as shown in Fig. 2a [25,101]. These organometallic compounds are easily detectable by ESI-HRMS as cations due to the efficient abstraction of the halogen anion. Following this approach, we collected the most abundant size- and termination-selected halopolyynes, namely $HC_{2n}X$ with n=4-5 from DCM and Cyhex/DBM, and $XC_{2n}X$ with n=4-5 from DCM, through HPLC, and we derivatized them with $Pd(PPh_3)_4$ as described in the Materials and Methods section. The monohalogenated compounds $HC_8X$ and $HC_{10}X$ gave expected mass spectra after derivatization (Figs. 2a-e). Ionization efficiency using an ESI ion source is high enough to record mass spectra even for these challenging, unstable, and low-concentration samples. Thanks to the palladium center, the distinctive isotope pattern confirmed halogen termination unambiguously. Cations derived from chloropolyynes were more visible due to their higher concentration compared to bromopolyynes. The strong agreement between experimental and simulated spectra definitely validates our assignment. Despite many attempts, we were not able to observe similar spectra for dihalogenated species $XC_{2n}X$. This can be explained by the significantly lower concentration of these species (see Figure S.2 in the SI) and probably lower stability in the ion source.

Although we demonstrated the formation of halopolyynes by PLAL, the absolute quantities produced remain modest. The production yield of polyynes by PLAL can be estimated in the order of $10^{-6}$ M $min^{-1}$, as observed in our previous work on H-capped polyynes [47]. Therefore, the material collected after each ablation is sufficient for analytical characterization (HPLC and HRMS) but remains orders of magnitude below the output of established wet chemical syntheses that deliver preparative quantities. Nevertheless, PLAL offers exceptional versatility, enabling the one-step production of a broad range of carbon atomic wires with different lengths and terminations, including relatively long chains, without the need for complex multistep chemical procedures.

### 3.2 Formation Mechanism of Halopolyynes

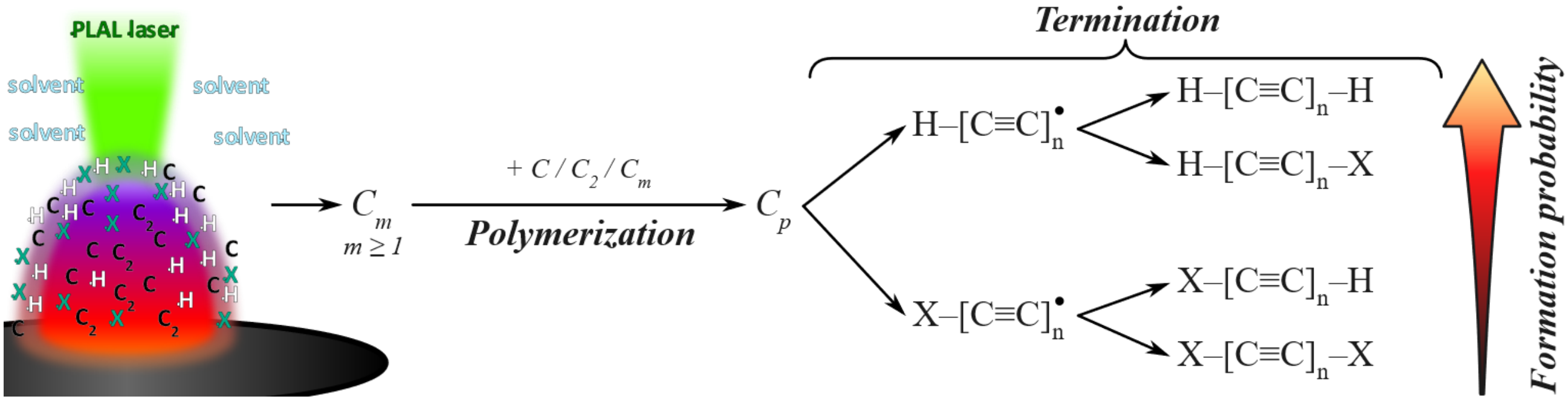

**Figure 3** Schematic of polyyne formation during PLAL in halogenated organic solvents. From left to right: the confined plasma plume produced by the laser ablation of a graphite target is shown, surrounded by solvent. Some carbon species, such as $C_m$ (m≥1), originate from target vaporization, while C, H, and X (halogen atom) result from solvent atomization. The schematic also illustrates the polymerization reactions and possible termination pathways, arranged in order of formation probability as inferred from HPLC measurements.

While the present data do not allow a direct determination of the underlying formation mechanism of halopolyynes during PLAL experiments, we propose a possible pathway here, inferred from established PLAL mechanisms reported in literature [7,30,44] and from the experimental trends observed in this work. This reaction follows a similar scheme to that of H-capped polyynes, i.e., a polymerization reaction followed by termination, and is sketched in Fig. 3. In PLAL, the high peak power density of the laser pulses locally vaporizes the graphite target, generating a variety of highly energetic carbon species (carbon atoms, dimers, radicals, and ions). These species are confined by the surrounding solvent, leading to the formation of a high-temperature, dense plasma plume characterized by steep temperature and pressure gradients, mainly localized at the boundary between the expanding plasma plume and the surrounding liquid [7,37,40,44,47,77,87]. The intense heat of the plasma plume induces the atomization of the nearby solvent molecules. As a result, the organic solvent acts as a secondary carbon source that, together with carbon species from the target, serves as the basis for the polymerization reactions, as discussed in other works [7,42,44,47,54]. In addition to C atoms, atomized halogenated solvent molecules are expected to generate reactive atomic species (H and X in Fig. 3) and hydrogen- and/or halogen- containing species, which can act as terminations for the growing carbon wire. Based on previous results, the most likely termination reaction products under these conditions are H-capped polyynes ($HC_{2n}H$), halogen-capped polyynes ($HC_{2n}X$), and dihalogen-capped polyynes ($XC_{2n}X$), as shown in Fig. 3.

We can gain insights into the dynamics and efficiency of these termination reactions by analyzing the yield of halopolyynes in contrast to the production of hydrogenated wires. We first ensured that the sample handling procedures do not significantly distort the relative composition of the polyyne mixture. The liquid-liquid phase extractions described in the Materials and Methods section efficiently transfer both H-capped and halogenated polyynes from the ablated halogenated organic solvent into acetonitrile, considering that chains of similar length exhibit comparable volatility and thermal degradation during rotary evaporation. As a result, the distribution of polyynes in acetonitrile remains consistent with the *as-prepared* one. Consequently, by comparing the chromatographic areas of each halopolyyne with H-capped polyynes of similar chain length, we obtain a ratio of their concentration, as these areas are directly related to compound concentration. Furthermore, this comparison relies on the assumption that hydrogenated polyynes, whose molar extinction coefficients are known, and halopolyynes of identical chain length, whose extinction coefficients are not experimentally available, have comparable absorptivities within the uncertainty required for qualitative trend analysis, rather than for absolute quantification. Within these limitations, the chromatographic area ratios provide a semi-quantitative description of the relative production yields of halopolyynes.

Halopolyynes' maximum yield is reached for Cl-capped chains; $HC_6Cl$, in particular, reaches 80% of $HC_6H$ concentration, as shown in Fig. S.2a in the SI. Br-capped polyynes are produced in much lower yield than Cl analogs (see Fig. S.2b in the SI). This difference is likely influenced by the composition of the ablation medium, which is not a pure halogenated solvent but a 2/3 v/v Cyhex/DBM mixture (see the Materials and Methods section). Under these conditions, the maximum yield is observed for $HC_{10}Br$, reaching 12% of $HC_{10}H$ concentration. Dihalogen-terminated wires show a much lower synthesis yield, as displayed in Fig. S.2c in the SI, ranging from 5 to 38 times lower than Cl-capped analogs for $ClC_{2n}Cl$, and from 79 to 140 times lower than Br-capped analogs for $BrC_{2n}Br$. A detailed list of relative synthesis yields of halopolyynes is presented in Table S.2 in the SI.

Despite their lower concentrations compared to hydrogenated chains, the observed relative yields indicate that the formation of halopolyynes is quite remarkable. Overall, the experimental evidence indicates that short, monohalogenated chains are favored under the present synthesis conditions, as halopolyynes' relative yields with respect to H-capped polyynes decrease with increasing chain length. This may suggest that halogen termination is more effective during the early stages of wire growth, while longer wires are progressively less likely to be terminated by halogens or recovered, possibly due to competing degradation and secondary reactions during PLAL. Indeed, similar degradation phenomena during PLAL have been previously observed and studied for hydrogen-capped polyynes [47,77]. These observed length-dependent yields support a scenario in which halogen termination predominantly occurs at early stages of wire polymerization.

Regarding the role of halogen termination in chain stabilization, this effect is highly dependent on the specific halogen and chain length. Diiodopolyynes ($IC_{2n}I$), for instance, exhibit significant stability and can be isolated as crystalline solids up to $IC_{10}I$, owing to the ability of iodine atoms to participate in stabilizing halogen bonds that reduce crosslinking reactions and enable co-crystal formation [63]. Chloro- or bromo-terminated polyynes are generally more stable than hydrogenated counterparts [102,103]. Species such as $ClC_4Cl$ and $BrC_4Br$ can indeed be isolated as solids, though their stability rapidly decreases with increasing chain length [102,103]. In contrast, H-capped polyynes, such as $HC_4H$, polymerize even at low temperatures [104]. For longer chains (n > 2), most Cl- and Br-terminated polyynes, including those synthesized in this work, remain stable only in solution and cannot be isolated as solids.

Beyond the observed and classified halopolyynes, other terminations may result from the recombination of C, H, and X (halogen) atoms in the plasma phase. These combinations can form small functional groups, such as $CH_nX_{3-n}$ (n=1,2), which could theoretically terminate carbon atomic wires, providing additional types of halopolyynes. A similar phenomenon was already observed with the detection of methyl-capped polyynes in water [44]. In our case, chromatograms of ablations performed in halogenated solvents reveal traces of other compounds not observed in ablations carried out in i-PrOH (Fig. 1 and Fig. S.1 and Fig. S.3 in the SI). These compounds exhibit the characteristic polyynic vibronic pattern but cannot be attributed to any known polyynes, nor to the $HC_{2n}X$ or $XC_{2n}X$ series. Due to their low formation yield, the concentrations of these additional halopolyynes are quite low, making their detection and further characterization challenging. Consequently, no clear trend in their retention times (reported in Fig. S.3 in the SI) could be established, nor was it possible to further investigate their chemical structures.

### 3.3 Spectroscopic Characterization of Halopolyynes

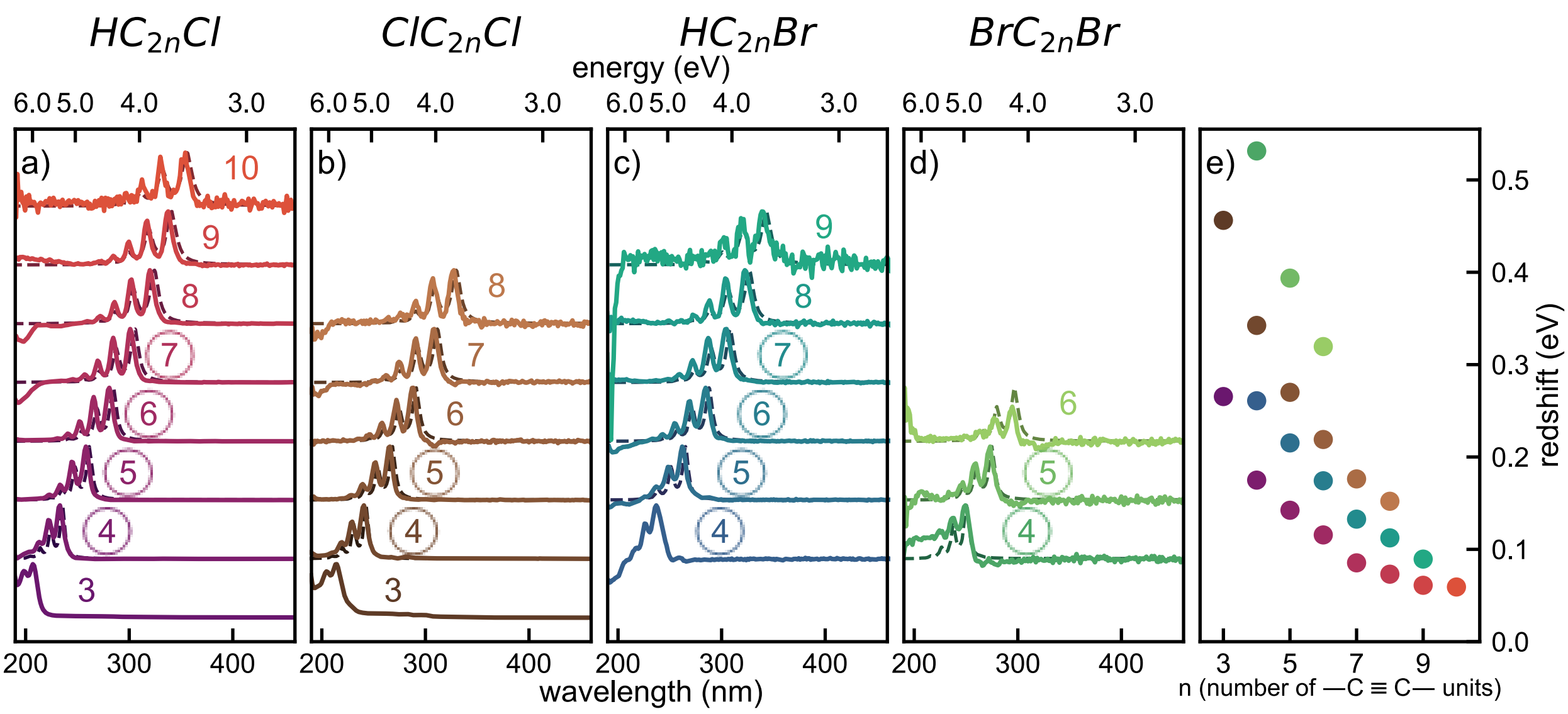

**Figure 4** Experimental UV-Vis spectra extracted from HPLC of isolated a) chloro- ($HC_{2n}Cl$, n=3-10), b) dichloro- ($ClC_{2n}Cl$, n=3-8), c) bromo- ($HC_{2n}Br$, n=4-9), and d) dibromo-capped ($BrC_{2n}Br$, n=4-6) polyynes. Solid colored lines represent experimental spectra, while dashed, darker lines report the available TD-DFT-calculated UV-Vis spectra. Chains selected for UV Resonance Raman spectroscopy are marked by circled labels. Numerical values of experimental vibronic transitions are listed in Table S.1 in the SI. e) Redshift of $E_{opt}$ for halopolyynes relative to their H-capped analogs of the same chain length, i.e., $E_{opt}(HC_{2n}H) - E_{opt}(YC_{2n}X)$ with Y=H, Cl, Br, and X=Cl, Br. Dots are color-coded to match panels a)-d).

Concentrated ablations (see the Materials and Methods section) enabled the separation and identification of a wide range of halopolyynes, whose UV-Vis absorption spectra extracted from the DAD of the HPLC system are displayed in Fig. 4. Correspondingly, we report the simulated UV-Vis spectra of halopolyynes, calculated using time-dependent density functional theory (TD-DFT) as described in the Materials and Methods section. The agreement between simulated and experimental spectra further confirms the correct assignment of each compound.

Halopolyynes feature the typical vibronic pattern seen in other polyynes, characterized by sharp, well-resolved absorption vibronic peaks corresponding to electronic transitions from the ground state to various vibrational levels of the first excited state. As with other conjugated systems, these vibronic series redshift with increasing chain length due to increased π-electron conjugation [4,41,44,105–108]. As previously mentioned, halogen termination induces a sizeable redshift compared to the corresponding hydrogenated analogs. This shift, quantified by the difference between the optical gaps of halopolyynes and H-capped chains, is reported in Fig. 4e. The effect is particularly prominent for Br-terminated and dihalogenated chains compared to Cl-terminated and monohalogenated counterparts, respectively. The stronger redshift in Br-capped polyynes reflects the more effective p-π conjugation of Br relative to Cl (see Fig. 4e), consistent with trends in auxochromic strength (I > Br > Cl) [96]. This enhanced p-π conjugation between halogen lone-pair electrons and the sp-carbon π-electron system promotes greater π-electron delocalization along the wire, extending the effective conjugation length and reducing bond length alternation (BLA), thus steering the structure towards a more cumulenic configuration (see Fig. S.4 in the SI). These structural changes manifest experimentally as a decrease in the bandgap, observed as a redshift in the vibronic progression.

A similar behavior is observed in dihalogenated chains, where the DFT-predicted BLA reduction (see Fig. S.4 in the SI) aligns with the more pronounced vibronic redshift shown in Fig. 4e. This demonstrates that dihalogen termination amplifies both structural and electronic effects compared to monohalogen termination.

We observe a slight reduction in the impact of halogen termination in longer halopolyynes, as evinced by the onset of saturation in the redshift in Fig. 4e. This trend arises from the diminishing size-confinement effect as the wires elongate and begin to approach the properties of ideal carbyne, although the chains analyzed here remain well below that regime [4,46].

From UV-Vis spectra, we also estimated the electron-phonon coupling in halopolyynes via the Huang-Rhys (HR) factor, as explained in our previous works [69,70,109]. The HR factors for monohalogenated polyynes are reported in Fig. S.5 in the SI and compared to H-capped polyynes of similar lengths [69]. Although the HR factors increase with chain length, halogen termination does not significantly enhance or suppress the electron-phonon coupling. This suggests that halogen termination, despite its non-negligible impact on the bandgap, exhibits a weaker auxochromic effect than endgroups containing sp-like bonds, such as CN, which strongly enhance conjugation and electron-phonon coupling [70]. These findings further indicate that the p-π conjugation introduced by halogen termination is structurally localized at the chain ends, with structural variations (i.e., BLA) primarily confined to the first triple bond (C≡C-X), in agreement with prior studies on related systems [110]. At the same time, its electronic effects propagate along the π-electron conjugated sp-carbon backbone.

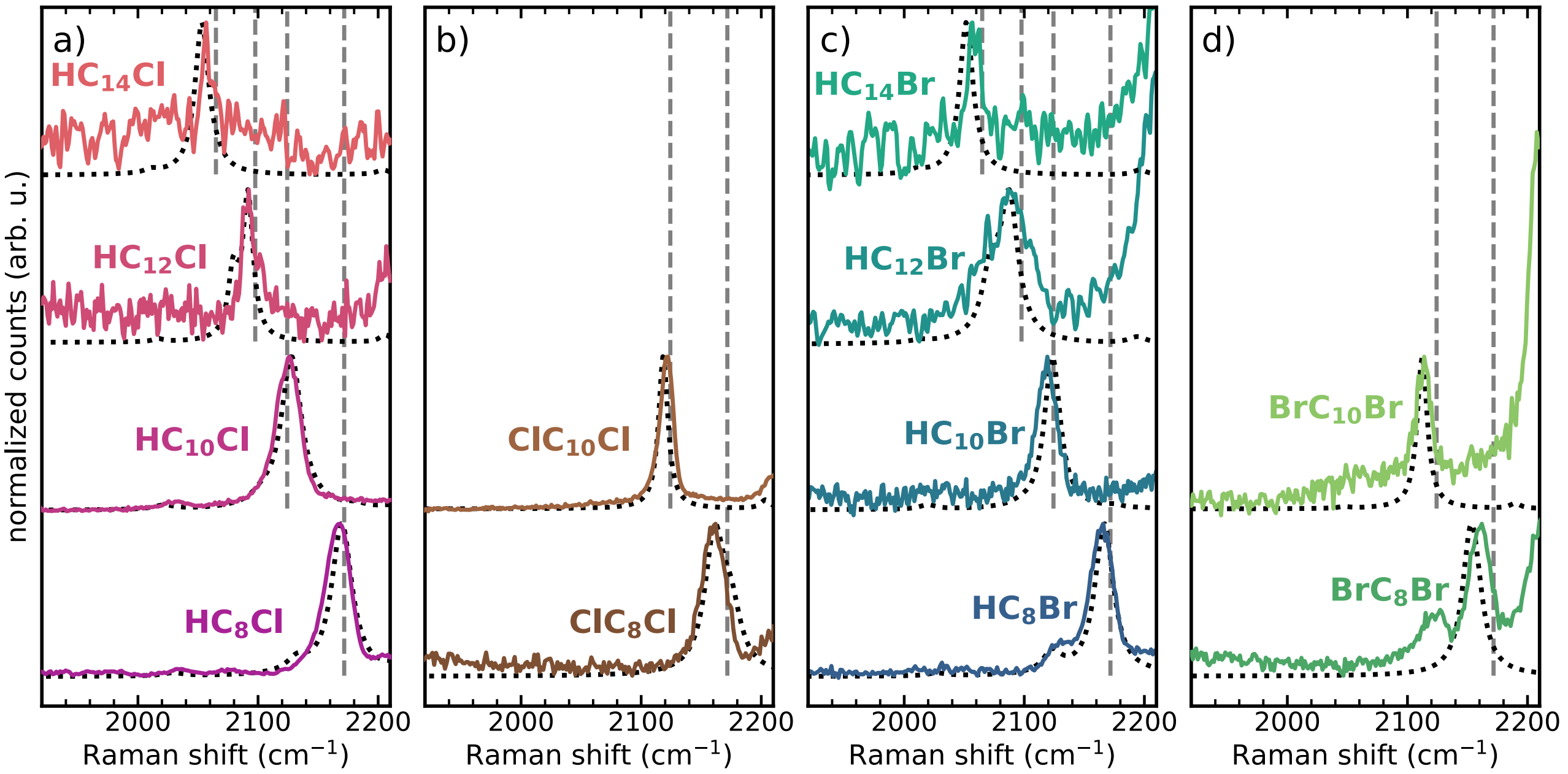

**Figure 5** Multi-wavelength UV Resonance Raman spectra of a) Cl-capped ($HC_{2n}Cl$, n=4-7), b) $Cl_2$-capped ($ClC_{2n}Cl$, n=4,5), c) Br-capped ($HC_{2n}Br$, n=4-7), and d) $Br_2$-capped ($BrC_{2n}Br$, n=4,5) polyynes. Black dashed lines represent DFT-simulated Raman spectra (frequency scaling factor 0.9407). Dashed gray lines indicate ECC mode frequencies of H-capped polyynes *[70]* of the same chain length. Measurement details and full Raman spectra are provided in Table S.3 and Fig. S.6, respectively, in the SI.

We employed UV Resonance Raman spectroscopy to characterize the vibrational properties of halopolyynes. The resonance condition enhances the Raman response by several orders of magnitude, enabling the detection of even low-concentrated, size- and termination-selected halopolyynes ($10^{-6}$ to $10^{-8}$ mol/L). Figure 5 presents UVRR spectra for the most concentrated halopolyynes collected via HPLC, for which resonance conditions could be matched, within the limitations of the synchrotron undulator gap aperture and optical transmission, at the IUVS beamline (Elettra synchrotron), which provides a tunable UV source from 200 to 272 nm [89]. Full experimental details and full UVRR spectra for each species are provided in Table S.3 and Fig. S.6, respectively, in the SI. Although the noise level in several spectra limits precise and thorough analysis, qualitative insights into the impact of the halogen terminations on the vibrational properties are still achievable.

Halopolyynes feature a strong and distinct Raman mode whose frequency downshifts with increasing chain length and different halogen terminations. Given its frequency range (from 2000 to 2200 $cm^{-1}$), sensitivity to chain length, and strong Raman activity, this mode is attributed to the collective stretching of the triple bonds in the sp-carbon chain, commonly referred to as the ECC mode (from the effective conjugation coordinate theory) or α mode [67,111]. DFT calculations, reported in Fig. 5, confirm this assignment. Frequency tuning with chain length is well-documented in prior studies [5,29,67,69,70,110,112–114].

Here, instead, we focus on the effect of halogen termination. Comparing the ECC mode frequencies of halopolyynes with H-capped polyynes of equal length, we observed a clear downshift in the ECC modes (refer to Table S.4 in the SI for the exact numerical values). This downshift in monohalogenated chains supports the presence of p-π conjugation between the halogen lone-pair electrons and the sp-carbon backbone, which extends the effective π-electron conjugation and reduces BLA (see Fig. S.4 in the SI), thus lowering the ECC mode frequency. This interpretation corroborates the redshifts observed in UV-Vis spectra.

For dihalogenated chains, we observe an even larger downshift, nearly double that of monohalogenated wires. The addition of a second halogen termination further reduces BLA values in these species (see Fig. S.4 in the SI), pushing their structures towards a more cumulenic configuration and justifying the observed downshift in the ECC mode.

Halogen termination also introduces additional Raman-active modes within the ECC mode frequency region. While some of these modes are detected experimentally, others are predicted by DFT but remain unobserved in our experiments (see Fig. 5 and Table S.4 in the SI). These secondary peaks arise from confinement effects, growing in number and decreasing in Raman activity with longer chains, a trend seen in other types of polyynes [70,97,110,112]. The nature of the termination significantly influences these modes, as larger vibrational amplitudes localize near the chain ends. Although these additional modes and their evolution with chain length and termination may offer valuable insights into the vibrational properties of halopolyynes, a detailed exploration lies beyond the scope of this work.

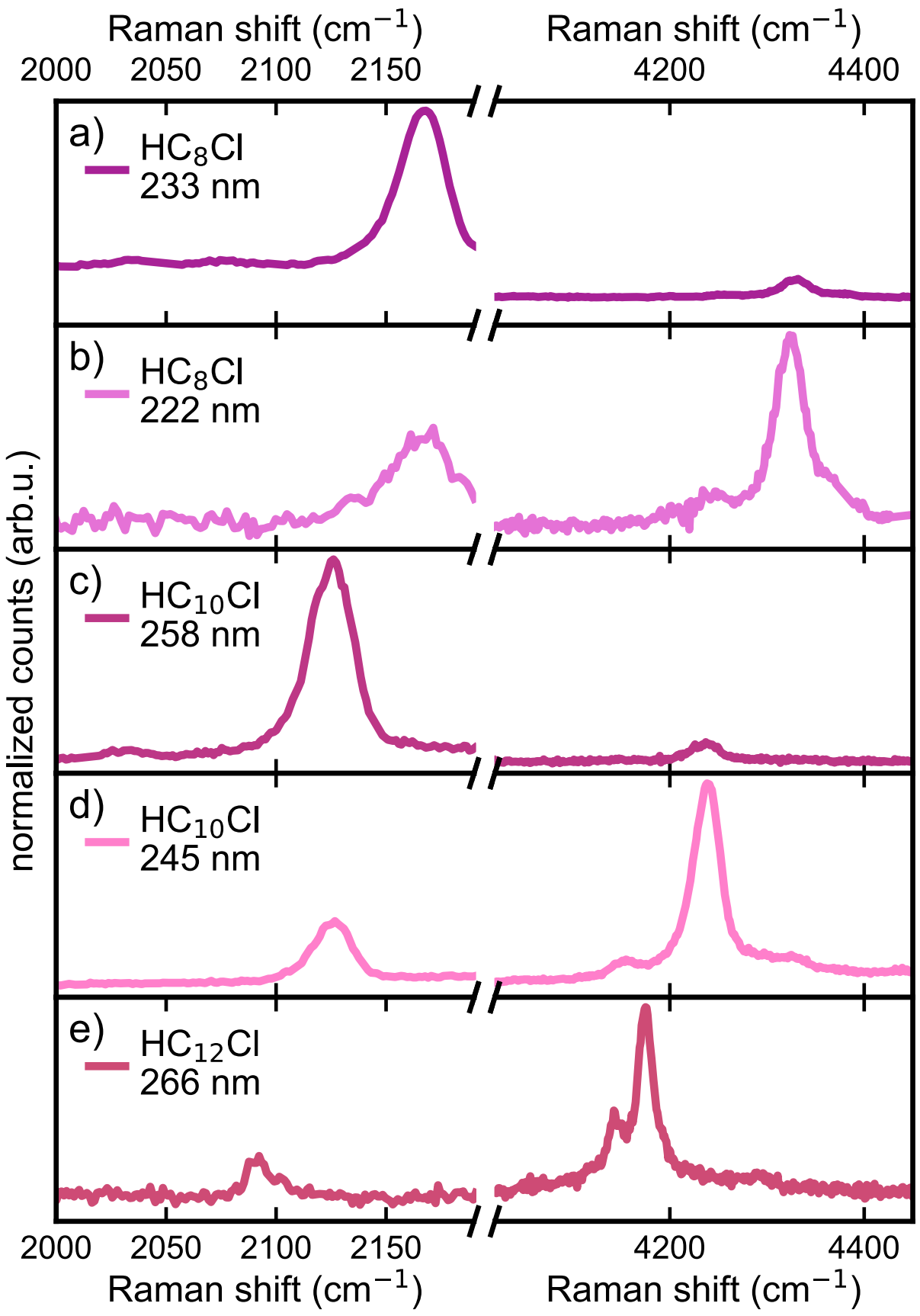

**Figure 6** UVRR spectra of $HC_8Cl$ excited at a) 233 nm ($|0\rangle_g \rightarrow |0\rangle_e$ transition) and b) 222 nm ($|0\rangle_g \rightarrow |1\rangle_e$ transition) in the fundamental and first overtone Raman shift regions. UVRR spectra of $HC_{10}Cl$ excited at c) 258 nm ($|0\rangle_g \rightarrow |0\rangle_e$ transition) and d) 245 nm ($|0\rangle_g \rightarrow |1\rangle_e$ transition) in the fundamental and first overtone Raman shift regions. e) UVRR spectrum of $HC_{12}Cl$ excited at 266 nm ($|0\rangle_g \rightarrow |1\rangle_e$ transition) in the fundamental and first overtone Raman shift regions. The $|0\rangle_g \rightarrow |0\rangle_e$ transition of $HC_{12}Cl$ at 281 nm could not be recorded due to limitations in the undulator gap. All spectra are normalized to their maximum intensity.

We also observed an enhancement of the second-order α (or ECC) mode (i.e., 2α mode) in Cl-capped polyynes compared to the fundamental α (or ECC) mode, particularly when the Raman excitation wavelength matches the electronic $|0\rangle_g \rightarrow |1\rangle_e$ transition (i.e., from the ground vibrational level of the ground electronic state, $|0\rangle_g$, to the first vibrational level of the first excited electronic state, $|1\rangle_e$). Supporting UVRR spectra are shown in Fig. 6. A similar phenomenon was reported for H-capped polyynes and explained using an analytical model derived from Albrecht's theory of resonance Raman scattering [69,115]. This enhancement results from a favorable overlap between the electronic wavefunctions involved in the Raman process, which boosts the 2α mode intensity relative to the α mode at this resonance condition.

These findings further evidence the strong and tunable electron-phonon coupling in halopolyynes and carbon atomic wires in general. Moreover, this overtone enhancement appears as a fingerprint of polyynes, as it is shared by several chains regardless of their terminations and lengths. Our observation suggests that this phenomenon could be a general characteristic of carbyne-like systems, though further studies are needed to confirm its universality. Investigations on long linear carbon chains encapsulated in carbon nanotubes could reveal whether this phenomenon is present in other realizations of carbyne and thus intrinsic to carbyne itself.

Finally, we extracted the vibrational anharmonicity parameter χ for halopolyynes $HC_8Cl$ and $HC_{10}Cl$, where multiple ECC mode overtones were observed (Fig. S.7 in the SI). Using the approach described in previous works [70,84], we obtained $\chi = 0.26 \cdot 10^{-2}$ for $HC_8Cl$ and $\chi = 0.37 \cdot 10^{-2}$ for $HC_{10}Cl$, in agreement with values reported for H-capped wires of similar lengths [70] and thus showing strong anharmonicity in halopolyynes. These results are consistent with the predictions of the universal law for the vibrational anharmonicity of carbyne-like materials, namely, $\chi = 0.28 \cdot 10^{-2}$ for $HC_8Cl$ and $\chi = 0.38 \cdot 10^{-2}$ for $HC_{10}Cl$ [84]. This provides the first independent confirmation of the vibrational anharmonicity universal law in a new class of carbon atomic wires, affirming halopolyynes as carbyne-like materials. Their well-defined, tunable structures and chemically selective halogen reactivity make them ideal platforms for exploring carbyne properties and engineering new sp-carbon-based molecular architectures.

## 4. Conclusions

In this work, we demonstrated a novel, one-step physical synthesis of halogen-terminated carbon atomic wires (halopolyynes) using pulsed laser ablation in liquid. Although previous PLAL works in halogenated solvents primarily yielded halogenated carbon nanoparticles [116,117], we successfully produced solutions of polydisperse monohalogenated ($HC_{2n}X$) and dihalogenated ($XC_{2n}X$) polyynes (X=Cl or Br, n=3-10), which were separated and analyzed by high-performance liquid chromatography. From this chromatographic data, we proposed a plausible formation mechanism in which highly reactive carbon species polymerize and are terminated by halogen atoms generated from solvent atomization during the plasma phase.

Through UV-Vis absorption and synchrotron-based UV Resonance Raman spectroscopy, we provided the first extensive spectroscopic characterization of halopolyynes. Halogen terminations act as auxochromes, participating in p-π conjugation with the sp-carbon backbone and inducing moderate redshifts in vibronic patterns and softening of collective Raman modes. Despite this effect on the electronic structure, the electron-phonon coupling remains comparable to that of H-capped wires, consistent with a conjugation effect that is electronically delocalized but structurally localized at the chain termination. Halopolyynes also show selective overtone enhancement and strong anharmonic behavior, consistent with other carbyne-like systems and marking the first independent confirmation of the universal anharmonicity law for carbyne-like materials [84].

Overall, this work establishes PLAL as a complementary platform to conventional wet chemistry. Beyond halopolyynes, the flexibility of PLAL opens synthetic pathways to explore new termination chemistries (e.g., thiol-capped polyynes or even sulfur-capped cumulenes) and carbon-rich architectures, expanding the toolbox for studying and engineering sp-carbon materials under non-equilibrium conditions.

## 5. Acknowledgments

P.M., S.M., S.P., A. C., P. S., and C.S.C. acknowledge funding from the European Research Council (ERC) under the European Union's Horizon 2020 research and innovation program ERC Consolidator Grant (ERC CoG2016 EspLORE grant agreement no. 724610, website: www.esplore.polimi.it). The authors acknowledge the CERIC-ERIC Consortium for access to experimental facilities and financial support (proposal numbers 20222105 and 20227204). P. P., B. P., and S. S. thank the National Science Centre Poland (Grant UMO-2022/45/B/ST4/01511) for support of this research. Calculations were carried out at the Wrocław Centre for Networking and Supercomputing (https://www.wcss.pl), Grant No. 523.

## 6. Data Availability Statement

The raw data required to reproduce the above findings are publicly available in the Zenodo repository at https://doi.org/10.5281/zenodo.15835613.

## 7. Declaration of generative AI and AI-assisted technologies in the manuscript preparation process

During the preparation of this work the author(s) used ChatGPT in order to rephrase and improve the flow of some sentences. After using this tool/service, the author(s) reviewed and edited the content as needed and take(s) full responsibility for the content of the published article.

# Supporting Information

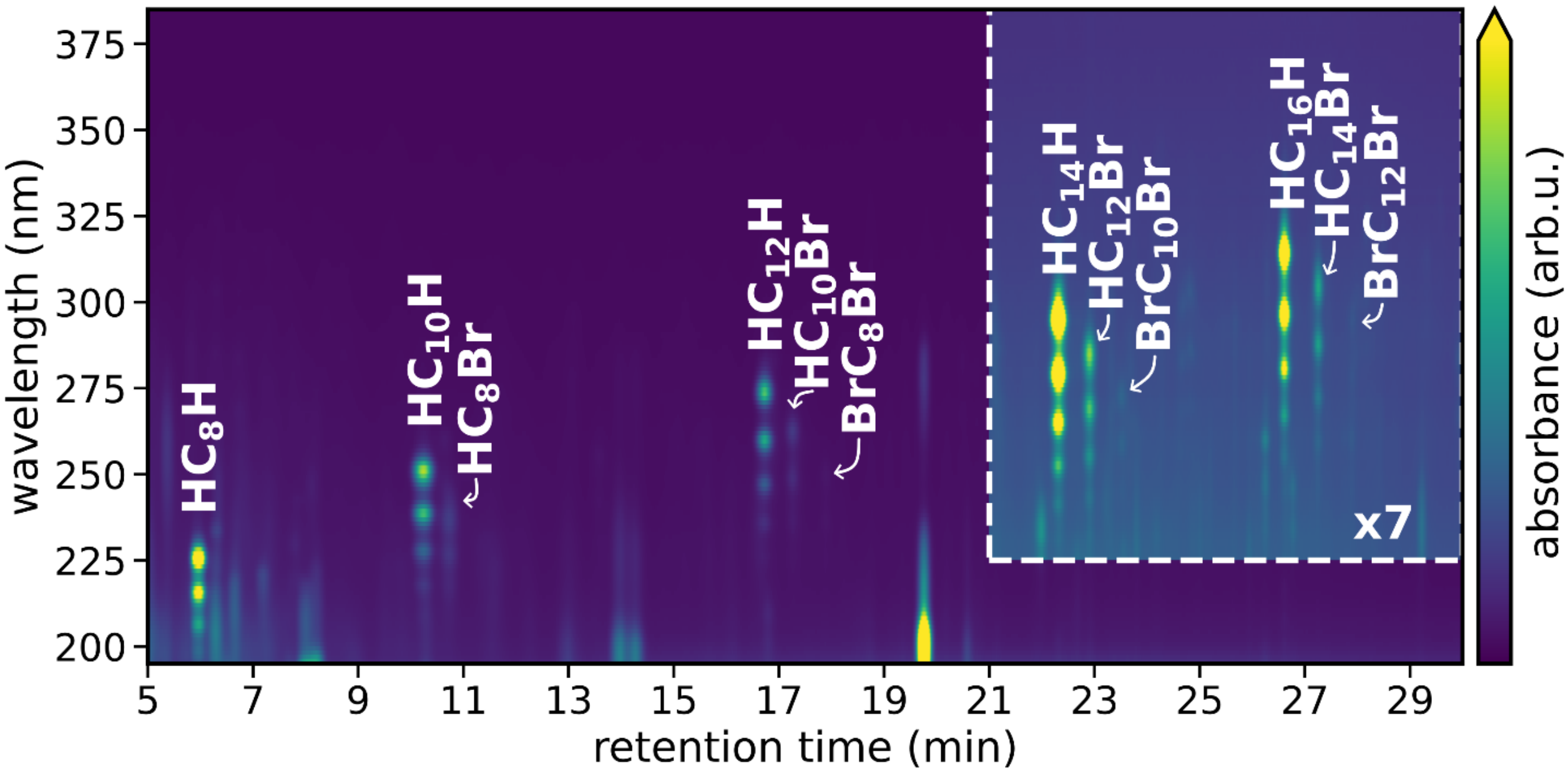

**Figure S.1** Portion of the HPLC analysis of an ablation in Cyhex/DBrM (after transfer in MeCN; see the Material and Methods section in the main text). This 2D representation shows the absorbance (color scale) as a function of the retention time and wavelength, extracted from the DAD coupled with the HPLC system.

| Polyyne | Ablation solvent | Retention time (min) | Vibronic peaks [nm] | Redshift from $HC_{2n}H$ (meV) |
|---|---|---|---|---|
| **$HC_6Cl$** | DCM | 6.5 | 207 199 | 241 |
| **$HC_8Cl$** | DCM | 11.5 | 233 222 213 | 165 |
| **$ClC_6Cl$** | DCM | 13.2 | 214 204 | 437 |
| **$HC_{10}Cl$** | DCM | 18.1 | 258 244 233 223 214 | 134 |
| **$ClC_8Cl$** | DCM | 19.5 | 241 229 219 | 341 |
| **$HC_{12}Cl$** | DCM | 23.7 | 281 266 252 241 231 222 | 129 |
| **$ClC_{10}Cl$** | DCM | 25.1 | 266 252 239 228 218 | 279 |
| **$HC_{14}Cl$** | DCM | 28.0 | 301 285 269 257 246 | 84 |
| **$ClC_{12}Cl$** | DCM | 29.5 | 288 272 258 244 | 237 |
| **$HC_{16}Cl$** | DCM | 32.6 | 320 302 286 272 259 | 74 |
| **$ClC_{14}Cl$** | DCM | 34.7 | 309 291 275 262 | 190 |
| **$HC_{18}Cl$** | DCM | 37.5 | 338 317 300 285 | 78 |
| **$ClC_{16}Cl$** | DCM | 38.2 | 328 307 291 276 | 169 |
| **$HC_{20}Cl$** | DCM | 41.2 | 352 330 312 | 40 |
| **$HC_8Br$** | Cyhex/DBrM | 10.5 | 237 226 218 | 255 |
| **$HC_{10}Br$** | Cyhex/DBrM | 16.9 | 262 248 237 227 | 207 |
| **$BrC_8Br$** | Cyhex/DBrM | 17.4 | 249 238 226 | 507 |
| **$HC_{12}Br$** | Cyhex/DBrM | 22.5 | 285 269 254 243 | 191 |
| **$BrC_{10}Br$** | Cyhex/DBrM | 23.1 | 273 259 246 | 398 |
| **$HC_{14}Br$** | Cyhex/DBrM | 26.9 | 305 287 272 259 248 | 138 |
| **$BrC_{12}Br$** | Cyhex/DBrM | 27.5 | 295 278 263 | 339 |
| **$HC_{16}Br$** | Cyhex/DBrM | 30.8 | 322 303 288 272 | 98 |
| **$HC_{18}Br$** | Cyhex/DBrM | 36.1 | 339 319 302 | 88 |

**Table S.1** Halopolyynes detected by HPLC from ablations of a graphite target in DCM or Cyhex/DBrM, both transferred in MeCN for the analysis, with the corresponding retention times from the chromatogram, the positions of the experimental UV-Vis absorption peaks, and the redshifts of the optical gap from the corresponding hydrogenated chain, calculated as $E_{opt}(HC_{2n}H) - E_{opt}(YC_{2n}X)$ where $E_{opt}$ is the equivalent energy of $\lambda_{max}$, Y=H, Cl, Br, and X=Cl, Br. The $\lambda_{max}$ of hydrogenated chains can be found in Table S.2 or in Ref. [1].

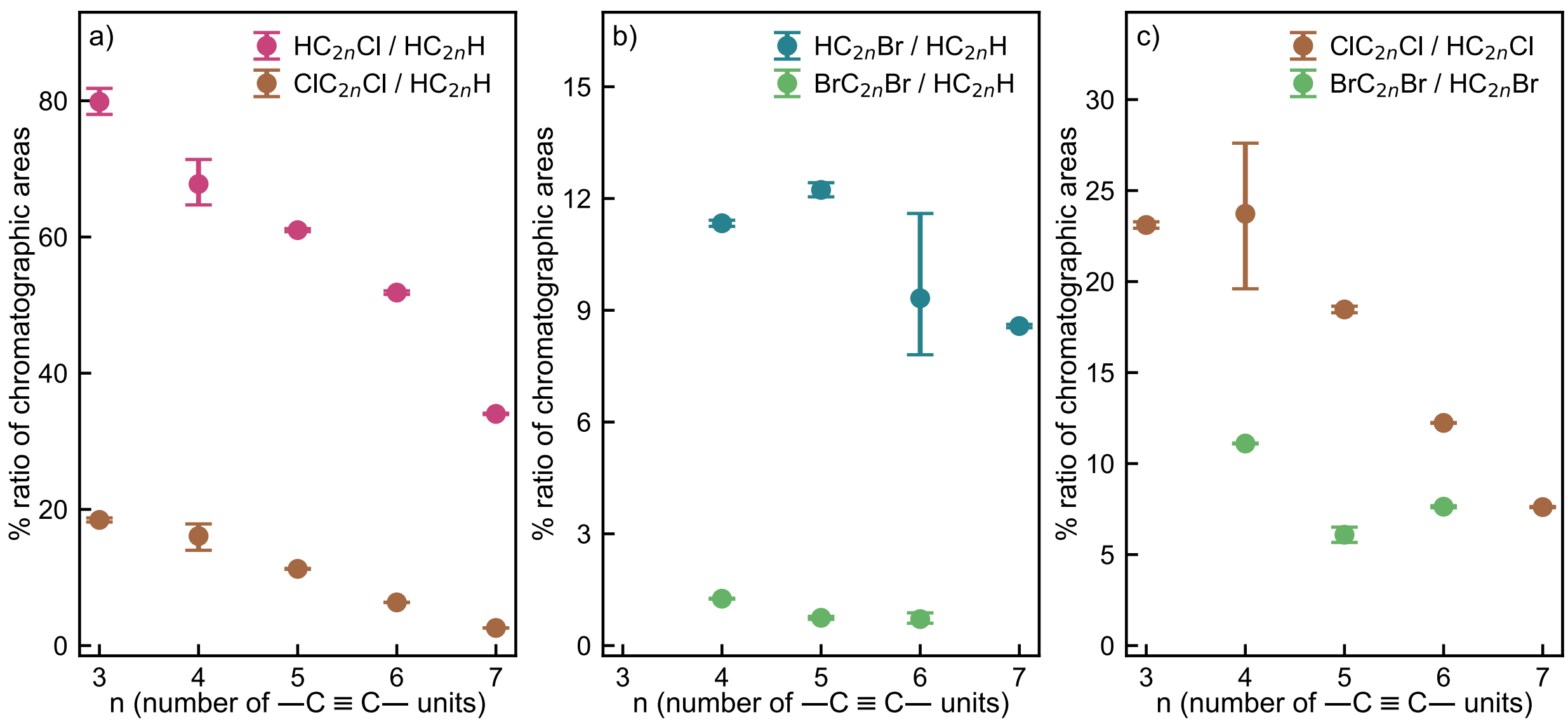

**Figure S.2** Percentage ratio of chromatographic areas of a) $HC_{2n}Cl$ and $ClC_{2n}Cl$ polyynes compared to $HC_{2n}H$ (n=3-7), b) $HC_{2n}Br$ (n=4-7) and $BrC_{2n}Br$ (n=4-6) compared to $HC_{2n}H$, and c) $ClC_{2n}Cl$ and $BrC_{2n}Br$ compared to $HC_{2n}Cl$ (n=3-7) and $HC_{2n}Br$ (n=4-6), respectively. Numerical values are reported in Table S.2.

| Halopolyyne | H-capped polyyne | Ratio (%) | Halopolyyne | H-capped polyyne | Ratio (%) |
|---|---|---|---|---|---|
| **$HC_6Cl$** (207) | $HC_6H$ (199) | 80 (78 – 82) | – | – | – |
| **$HC_8Cl$** (233) | $HC_8H$ (226) | 68 (65 – 71) | **$HC_8Br$** (237) | $HC_8H$ (226) | 11.34 (11.25 – 11.42) |
| **$HC_{10}Cl$** (258) | $HC_{10}H$ (251) | 61.0 (60.8 – 61.2) | **$HC_{10}Br$** (262) | $HC_{10}H$ (251) | 12.2 (12.1 – 12.4) |
| **$HC_{12}Cl$** (281) | $HC_{12}H$ (273) | 51.8 (51.6 – 52.1) | **$HC_{12}Br$** (285) | $HC_{12}H$ (273) | 9 (8 – 12) |
| **$HC_{14}Cl$** (301) | $HC_{14}H$ (295) | 34.0 (33.9 – 34.1) | **$HC_{14}Br$** (305) | $HC_{14}H$ (295) | 8.58 (8.53 – 8.62) |
| **$ClC_6Cl$** (214) | $HC_6H$ (199) | 18.5 (18.2 – 18.8) | – | – | – |
| **$ClC_8Cl$** (241) | $HC_8H$ (226) | 16 (14 – 18) | **$BrC_8Br$** (249) | $HC_8H$ (226) | 1.26 (1.25 -1.27) |
| **$ClC_{10}Cl$** (266) | $HC_{10}H$ (251) | 11.27 (11.20 – 11.33) | **$BrC_{10}Br$** (273) | $HC_{10}H$ (251) | 0.74 (0.70 – 0.78) |
| **$ClC_{12}Cl$** (288) | $HC_{12}H$ (273) | 6.34 (6.32 – 6.36) | **$BrC_{12}Br$** (295) | $HC_{12}H$ (273) | 0.7 (0.6 – 0.9) |
| **$ClC_{14}Cl$** (305) | $HC_{14}H$ (295) | 2.589 (2.586 – 2.592) | – | – | – |
| **Dihalopolyyne** | **Halopolyyne** | **Ratio (%)** | **Dihalopolyyne** | **Halopolyyne** | **Ratio (%)** |
| **$ClC_6Cl$** (214) | $HC_6Cl$ (207) | 23.1 (22.9 – 23.3) | – | – | – |
| **$ClC_8Cl$** (241) | $HC_8Cl$ (233) | 24 (20 – 28) | **$BrC_8Br$** (249) | $HC_8Br$ (237) | 11.10 (11.08 – 11.12) |
| **$ClC_{10}Cl$** (266) | $HC_{10}Cl$ (258) | 18.5 (18.3 -18.6) | **$BrC_{10}Br$** (273) | $HC_{10}Br$ (262) | 6.1 (5.7 – 6.5) |
| **$ClC_{12}Cl$** (288) | $HC_{12}Cl$ (281) | 12.24 (12.22 – 12.25) | **$BrC_{12}Br$** (295) | $HC_{12}Br$ (285) | 7.63 (7.57 – 7.69) |
| **$ClC_{14}Cl$** (305) | $HC_{14}Cl$ (301) | 7.61 (7.58 – 7.64) | – | – | – |

**Table S.2** Percentage ratio of chromatographic areas for halopolyynes compared to H-capped polyynes, and monohalogenated and dihalogenated wires. The number in parentheses next to each polyyne label indicates the chromatographic channel used for integration, corresponding to the most intense vibronic peak in the absorption spectrum of each wire. The numbers in parentheses within the ratio columns represent the lower and upper bounds of the percentage ratio, accounting for errors in fitting the chromatographic peaks.

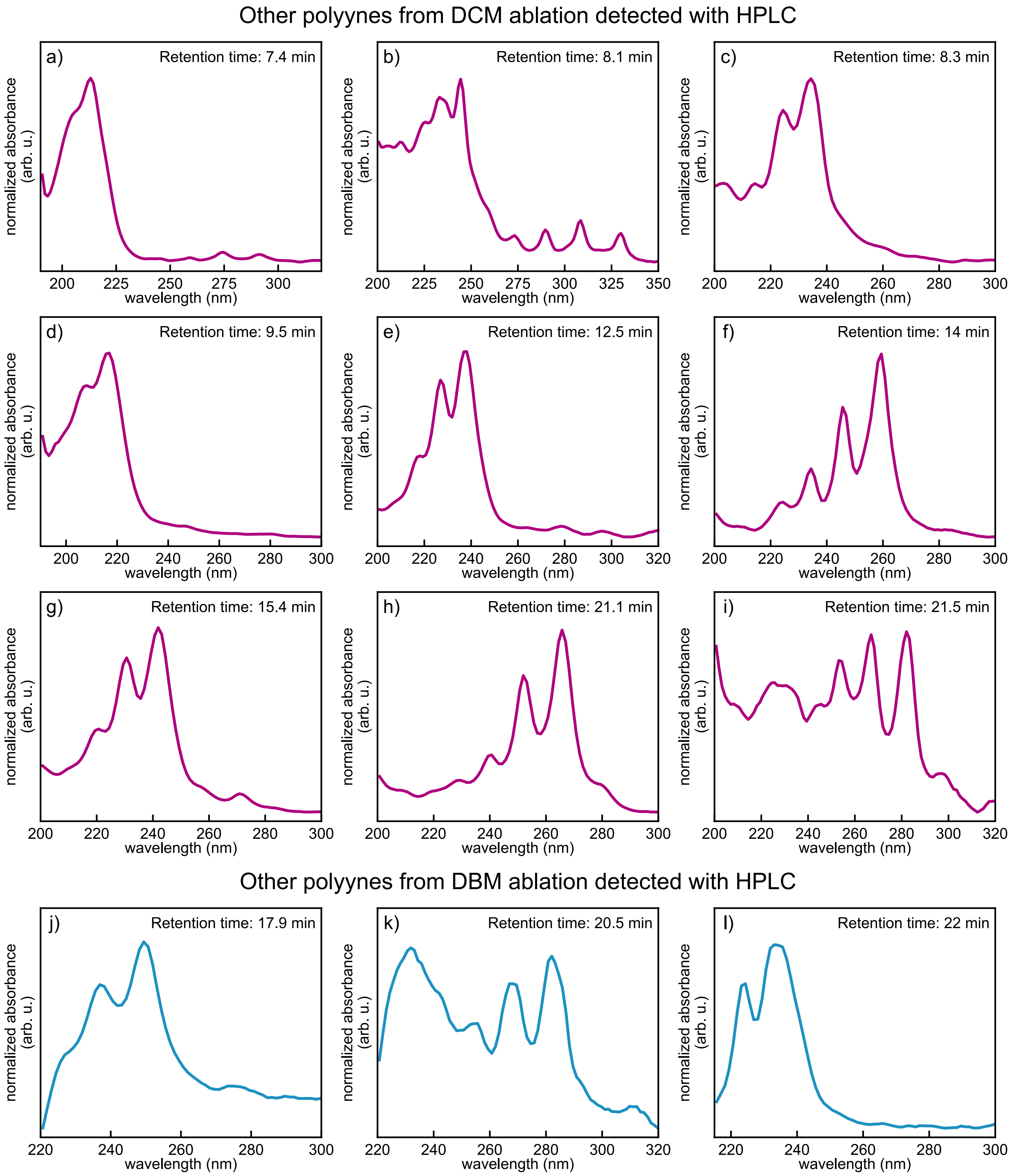

**Figure S.3** Experimental UV-Vis spectra extracted from HPLC for isolated unknown compounds detected from ablations in a–i) DCM or j–l) DBM. The absorbance in each spectrum is normalized to its maximum within the displayed range. The retention time for each spectrum is indicated in the top right corner of each panel.

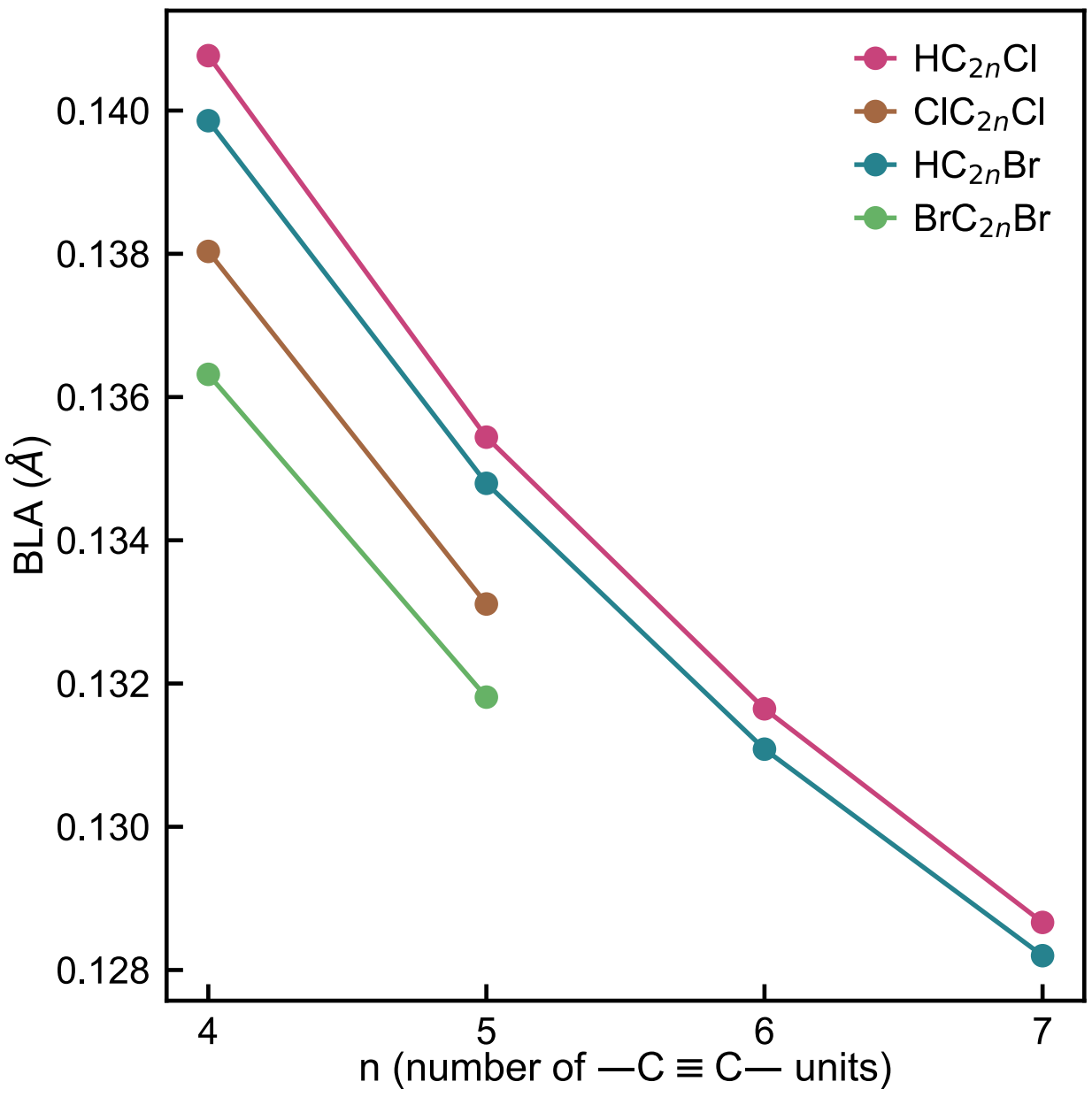

**Figure S.4** Bond length alternation (BLA) values for halopolyynes ($HC_{2n}Cl$, n=4-7, in pink; $ClC_{2n}Cl$, n=4,5, in brown; $HC_{2n}Br$, n=4-7, in blue; $BrC_{2n}Br$, n=4,5, in green) calculated via DFT.

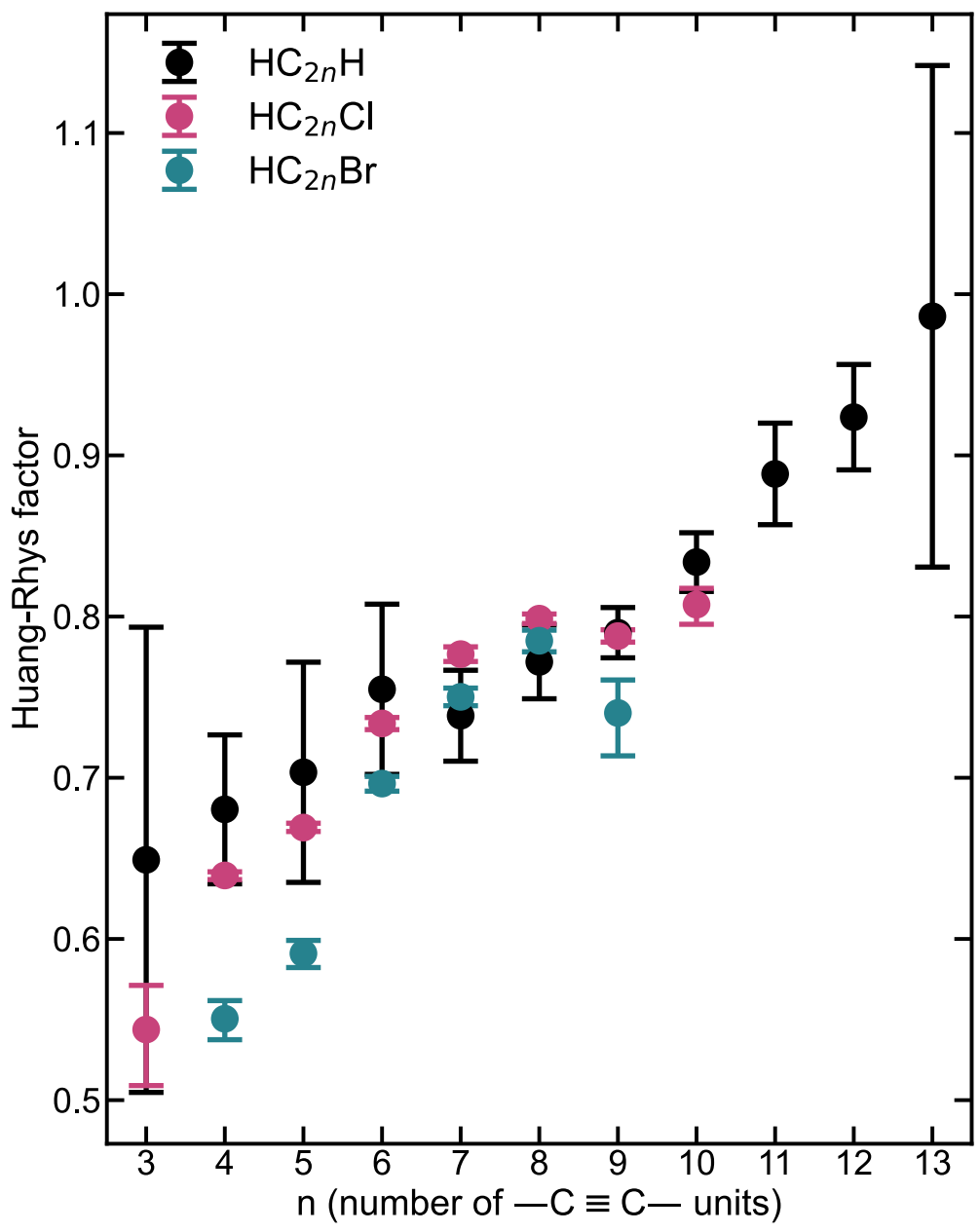

**Figure S.5** Huang-Rhys factor for hydrogen- (black circles) [2], Cl- (pinkish circles), and Br-capped (blueish circles) polyynes, extracted from the UV-Vis spectra of Fig. 4 in the main text. The error bars are due to the error of the model used to fit the UV-Vis absorption spectra.

| Polyyne | Solvent | Excitation wavelength [nm] | | Figure |
|---|---|---|---|---|
| **$HC_8Cl$** | Acetonitrile/water | 233 | (1) | Figs. 5 and 6 |
| | | 222 | (2) | Fig. 6, Fig. S.6 |
| **$HC_{10}Cl$** | Acetonitrile/water | 258 | (1) | Figs. 5 and 6 |
| | | 245 | (2) | Fig. 6, Fig. S.6 |
| **$HC_{12}Cl$** | Acetonitrile/water | 266 | (2) | Figs. 5 and 6 |
| **$HC_{14}Cl$** | Acetonitrile/water | 269 | (3) | Fig. 5 |
| **$ClC_8Cl$** | Acetonitrile/water | 240 | (1) | Fig. 5 |
| **$ClC_{10}Cl$** | Acetonitrile/water | 266 | (1) | Fig. 5 |
| **$HC_8Br$** | Acetonitrile/water | 237 | (1) | Fig. 5 |
| **$HC_{10}Br$** | Acetonitrile/water | 261 | (1) | Fig. 5 |
| **$HC_{12}Br$** | Acetonitrile/water | 269 | (2) | Fig. 5 |
| **$HC_{14}Br$** | Acetonitrile/water | 272 | (3) | Fig. 5 |
| **$BrC_8Br$** | Acetonitrile/water | 251 | ≈(1) | Fig. 5 |
| **$BrC_{10}Br$** | Acetonitrile/water | 266 | In between (1) and (2) | Fig. 5 |

**Table S.3** Solvent and synchrotron-based excitation wavelengths at which the Raman spectra in Figs. 5 and 6 and Fig. S.6 are recorded. In the parenthesis, we have indicated the vibronic transition at which we excited each specific wire ((1) means the first vibronic peak, refer to Table S.1). The reference figure in which each spectrum is indicated as well.

| Halopolyyne | DFT Raman frequency [cm$^{-1}$] | DFT Raman activity | Experimental Raman frequency [cm$^{-1}$] | Halopolyyne | DFT Raman frequency [cm$^{-1}$] | DFT Raman activity | Experimental Raman frequency [cm$^{-1}$] |
|---|---|---|---|---|---|---|---|
| **$HC_8Cl$** | 2029.37 | 0.0203 | 2032 | **$HC_8Br$** | 2027.31 | 0.0190 | 2028 |
| | 2132.61 | 0.0846 | — | | 2123.32 | 0.1397 | — |
| | 2169.77 (α mode) | 1 | 2163 | | 2166.51 (α mode) | 1 | 2159 |
| | 2211.94 | 0.0439 | — | | 2202.37 | 0.0107 | — |
| **$HC_{10}Cl$** | 2021.77 | 0.0269 | 2027 | **$HC_{10}Br$** | 2019.73 | 0.0282 | — |
| | 2103.39 | 0.0489 | — | | 2096.52 | 0.0125 | — |
| | 2127.85 (α mode) | 1 | 2121 | | 2123.38 (α mode) | 1 | 2118 |
| | 2181.60 | 0.0218 | — | | 2175.40 | 0.0168 | — |
| | 2208.97 | 0.0177 | — | | 2200.15 | 0.0049 | — |
| **$HC_{12}Cl$** | 2016.78 | 0.0210 | — | **$HC_{12}Br$** | 2014.12 | 0.0154 | — |
| | 2079.46 | 0.4594 | — | | 2072.33 | 0.4469 | — |
| | 2091.64 (α mode) | 1 | 2087 | | 2087.50 (α mode) | 1 | 2082 |
| | 2156.78 | 0.0002 | — | | 2181.04 | 0.0001 | — |
| | 2184.61 | 0.0011 | — | | 2194.33 | 0.0473 | — |
| | 2205.75 | 0.0531 | — | | | | — |
| **$HC_{14}Cl$** | 2011.36 | 0.0241 | — | **$HC_{14}Br$** | 2011.09 | 0.0272 | |
| | 2053.22 (α mode) | 1 | 2054 | | 2051.67 (α mode) | 1 | 2055 |
| | 2065.93 | 0.0407 | — | | 2062.48 | 0.0269 | — |
| | 2132.67 | 0.0002 | — | | 2128.11 | 0.0002 | — |
| | 2161.22 | 0.0003 | — | | 2159.83 | 0.0007 | — |
| | 2182.50 | 0.0041 | — | | 2179.13 | 0.0002 | — |
| | 2204.16 | 0.0333 | — | | 2196.29 | 0.0340 | — |
| **$ClC_8Cl$** | 2161.08 (α mode) | 1 | 2160 | **$BrC_8Br$** | 2152.50 (α mode) | 1 | 2157 |
| | 2174.82 | 0.4011 | — | | 2157.17 | 0.3106 | — |
| **$ClC_{10}Cl$** | 2055.18 | 0.0030 | — | **$BrC_{10}Br$** | 2047.32 | 0.01163 | — |
| | 2118.41 (α mode) | 1 | 2118 | | 2112.31 (α mode) | 1 | 2112 |
| | 2203.48 | 0.0565 | — | | 2189.78 | 0.0344 | — |

**Table S.4** DFT-calculated Raman frequencies and normalized activities for all Raman-active modes involving collective vibrations of the sp-carbon backbone in halopolyynes. Experimental frequencies are provided for modes observed in the Raman spectra shown in Fig. 5 of the main text. For unobserved modes, a “—” symbol is indicated in the corresponding cell.

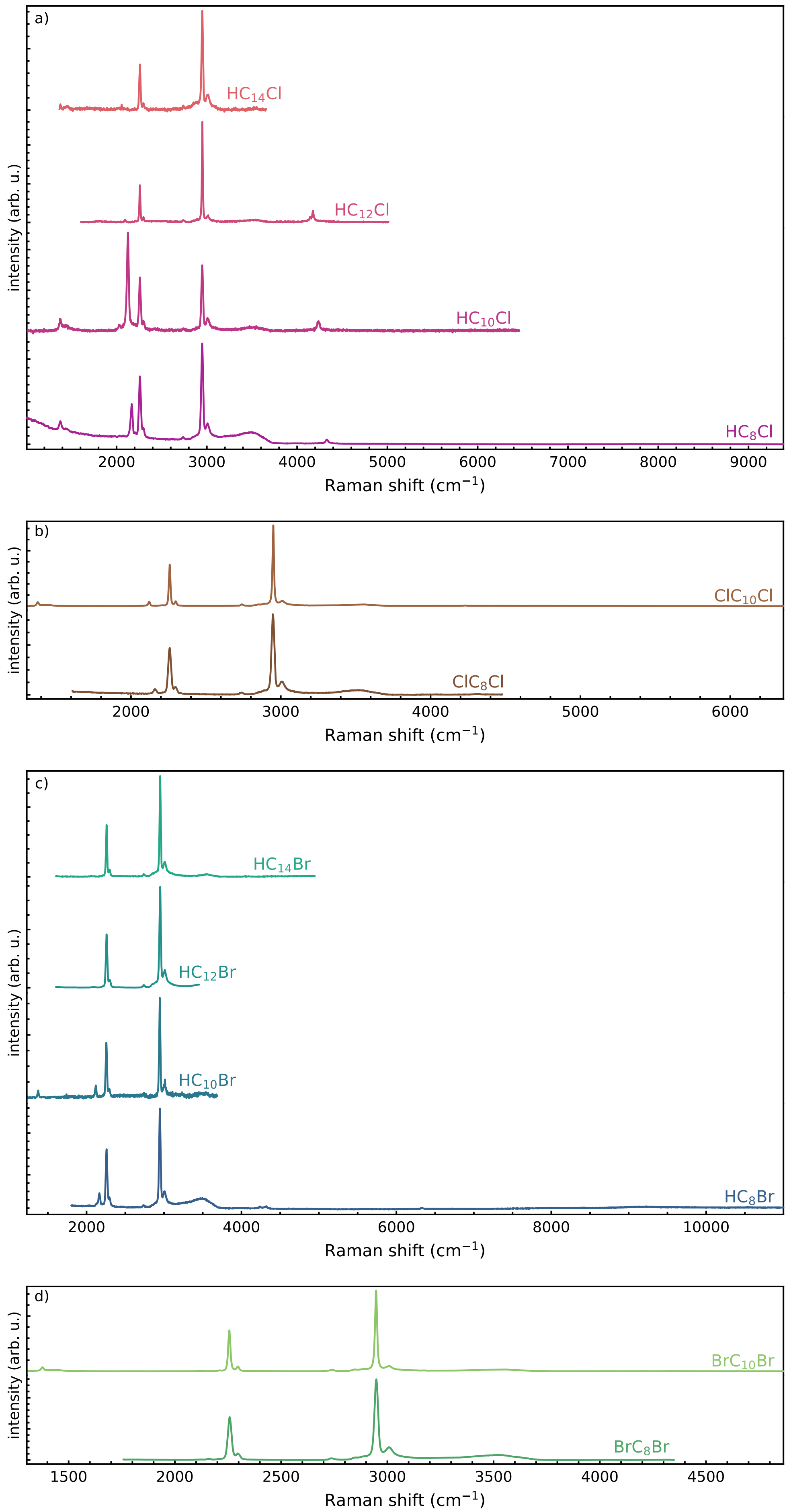

a)
HC14Cl
HC12Cl
HC10Cl
HC8Cl
intensity (arb. u.)
Raman shift (cm−1)
b)
ClC10Cl
ClC8Cl
intensity (arb. u.)
Raman shift (cm−1)
c)
HC14Br
HC12Br
HC10Br
HC8Br
intensity (arb. u.)
Raman shift (cm−1)
d)
BrC10Br
BrC8Br
intensity (arb. u.)
Raman shift (cm−1)

**Figure S.6** Multi-wavelength UV Resonance Raman spectra of a) Cl-capped ($HC_{2n}Cl$, n=4-7), b) $Cl_2$-capped ($ClC_{2n}Cl$, n=4,5), c) Br-capped ($HC_{2n}Br$, n=4-7), and d) $Br_2$-capped ($BrC_{2n}Br$, n=4,5) polyynes. Measurements details are provided in Table S.3 in the SI.

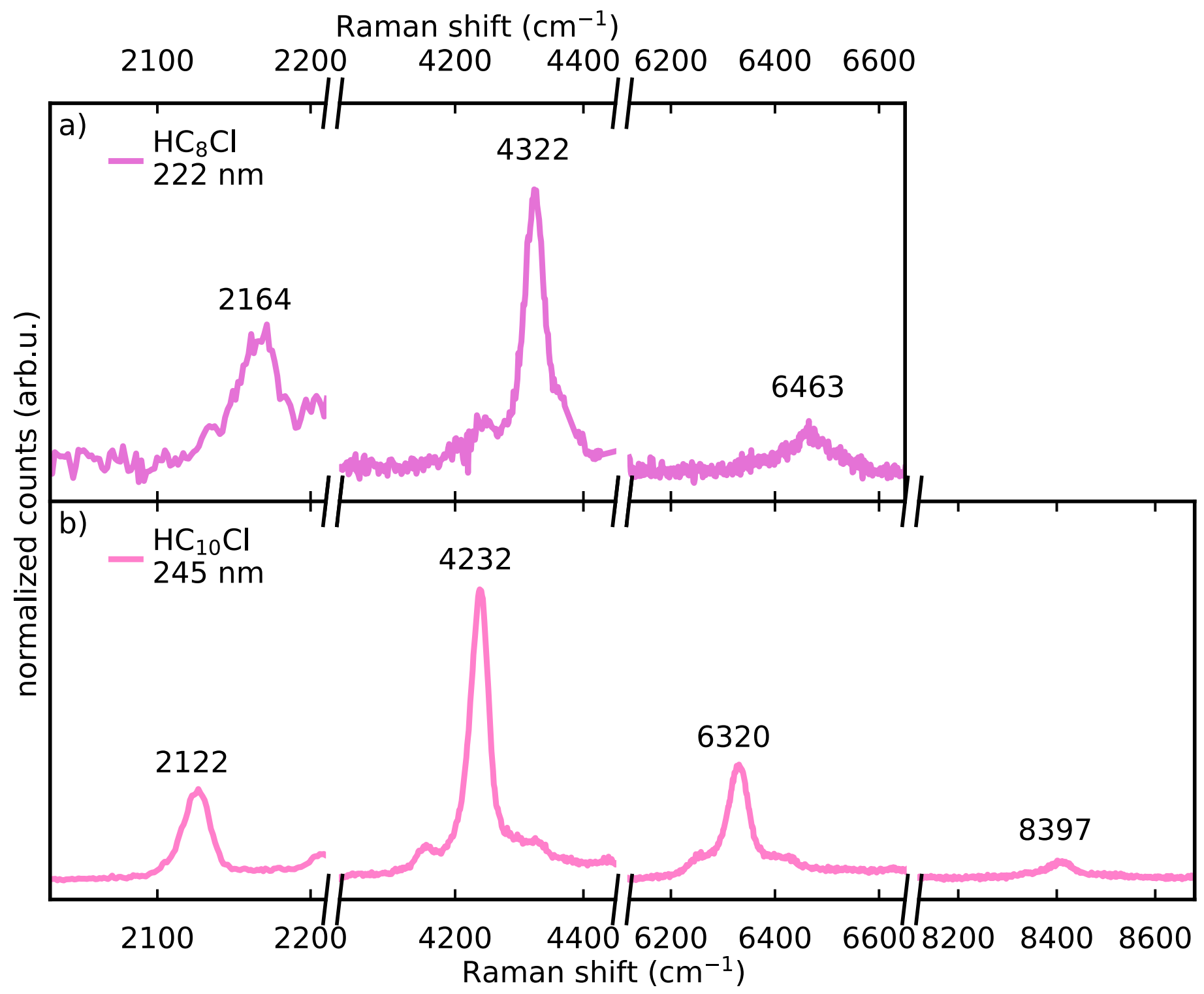

**Figure S.7** UV resonance Raman spectra of a) $HC_8Cl$ (excited at 222 nm) and $HC_{10}Cl$ (excited at 245 nm) excited at their corresponding $|0\rangle_g \rightarrow |1\rangle_e$ transitions, showing their ECC modes and their overtones. The spectra have been calibrated using the CN stretching mode of acetonitrile (not shown in the spectra). The frequency of the ECC mode and its overtones are reported in each panel accordingly.